\documentclass{emulateapj}
\usepackage{times}
\newcommand{\D}[2]{\frac{\partial #2}{\partial #1}}
\newcommand\bb[1]{\mbox{\boldmath{$#1$}}}
\newcommand\del{\bb{\nabla}} 
\newcommand\bcdot{\bb{\cdot}}
\newcommand\btimes{\bb{\times}}

\begin{document}
%\submitted{Submitted to the Astrophysical Journal, \today}
%\submitted{Draft version, \today}

\shorttitle{\textsc{Viscous, Resistive Magnetorotational Modes}}
\shortauthors{\textsc{Pessah and Chan}}
 
\title{\textsc{Viscous, Resistive Magnetorotational Modes}}

\author{Martin E. Pessah}
\affil{School of Natural Sciences, Institute for Advanced Study, Princeton, NJ, 08540}
\and
\author{Chi-kwan Chan}
\affil{Institute for Theory and Computation,
Harvard-Smithsonian Center for Astrophysics,
60 Garden Street, Cambridge, MA 02138}
\email{mpessah@ias.edu, ckchan@cfa.harvard.edu}

\begin{abstract}
  We carry out a comprehensive analysis of the behavior of the
  magnetorotational instability (MRI) in viscous, resistive plasmas.
  We find exact, non-linear solutions of the non-ideal
  magnetohydrodynamic (MHD) equations describing the local dynamics of
  an incompressible, differentially rotating background threaded by a
  vertical magnetic field when disturbances with wavenumbers
  perpendicular to the shear are considered. We provide a geometrical
  description of these viscous, resistive MRI modes and show how their
  physical structure is modified as a function of the Reynolds and
  magnetic Reynolds numbers. We demonstrate that when finite
  dissipative effects are considered, velocity and magnetic field
  disturbances are no longer orthogonal (as it is the case in the
  ideal MHD limit) unless the magnetic Prandtl number is unity.  We
  generalize previous results found in the ideal limit and show that a
  series of key properties of the mean Reynolds and Maxwell stresses
  also hold for the viscous, resistive MRI. In particular, we show
  that the Reynolds stress is always positive and the Maxwell stress
  is always negative.  Therefore, even in the presence of viscosity
  and resistivity, the total mean angular momentum transport is always
  directed outwards.  We also find that, for any combination of the
  Reynolds and magnetic Reynolds numbers, magnetic disturbances
  dominate both the energetics and the transport of angular momentum
  and that the total mean energy density is an upper bound for the
  total mean stress responsible for angular momentum transport. The
  ratios between the Maxwell and Reynolds stresses and between
  magnetic and kinetic energy densities increase with decreasing
  Reynolds numbers for any magnetic Reynolds number; the lowest limit
  of both ratios is reached in the ideal MHD regime.
\end{abstract}


\section{Introduction}
\label{sec:intro}

The magnetorotational instability \citep[MRI,][]{BH91, BH98} has been
widely studied in the inviscid and perfectly conducting,
magnetohydrodynamic (MHD) limit.  The departures from this idealized
situation are usually parametrized according to the Reynolds ${\rm Re}
= vl/\nu$ and magnetic Reynolds ${\rm Rm} = vl/\eta$ numbers, where
$v$ and $l$ stand for the relevant characteristic velocity and
lengthscale and $\nu$ and $\eta$ stand for the kinematic viscosity and
resistivity. The ideal MHD regime is then formally identified with the
limit ${\rm Re}~,~{\rm Rm}~\rightarrow~\infty$. There are many
situations of interest in which the effects of dissipation need to be
considered.

From the astrophysical point of view, accretion disks around young
stellar objects constitute one of the most compelling reasons for
investigating the MRI beyond the ideal limit. In particular, there is
great interest in understanding to what extent can MHD turbulence
driven by the MRI enable efficient angular momentum transport in cool,
poorly conducting, protoplanetary disks \citep[see, e.g.][]{BB94,
  Jin96, Gammie96, SM99, SW05}.  Most of the studies addressing the
effects of dissipation in non-ideal MRI have usually focused in
inviscid, resistive plasmas. However, accretion disks are
characterized by a wide range of magnetic Prandtl numbers, with ${\rm
  Pm}=\nu/\eta$ varying by several orders of magnitude across the
entire disk \citep[see, e.g.,][]{BH08}. In order to understand the
behavior of the MRI under these conditions it is necessary to relax
the assumption of an inviscid plasma.

A large fraction of the shearing box simulations addressing the
non-linear regime of the MRI have been carried out in the ideal MHD
limit, i.e., without including explicit dissipation in the codes
\citep[see, e.g.,][]{HGB95,Brandenburg95,Sanoetal04}.  However, even
in the absence of explicit viscosity and resistivity, finite
difference discretization leads to numerical diffusion/dissipation.
Therefore, even in this type of simulations, it is necessary to
understand the impact of these numerical artifacts that lead to
departures from the ideal MHD regime and how similar they are when
compared with physical (resolved) dissipation.

A handful of numerical studies with explicit resistivity but zero
physical viscosity have been carried out in order to understand the
effects of ohmic dissipation in the saturation of MRI-driven
turbulence \citep[see, e.g.,][]{SIM98, SI01, FSH00, Sanoetal04,
  TSD07}. In particular, \citet{SS03} have shown that the saturation
level of the stresses increases with increasing magnetic Reynolds
number and seem to converge to an asymptotic value for magnetic
Reynolds numbers larger than unity.

Recent work has pointed out problems with convergence in zero-net-flux
numerical simulations of ideal MHD driven by the MRI \citep{PCP07,
  FPI07} implying the necessity of incorporating explicit dissipation
in the codes.  Numerical studies with both resistivity and viscosity,
in the presence of a mean vertical magnetic field \citep{LL07} and in
the case of zero net flux \citep{FPII07}, have begun to uncover how
the characteristics of fully developed MRI-driven turbulence depends
on the Reynolds and magnetic Reynolds numbers.  Even though the ranges
in Reynolds and magnetic Reynolds numbers that can be currently
addressed is still limited, the results obtained from the simulations
suggest that the magnetic and kinetic energies contained in turbulent
motions in the saturated regime depend on the values of the
microphysical viscosity and resistivity. In particular, the mean
angular momentum transport in the turbulent state increases with
increasing magnetic Prandtl number.

From the experimental perspective, understanding the effects of
non-vanishing resistivity and viscosity in the behavior of the MRI
seems imperative, since the physical conditions achievable in the
laboratory depart significantly from the ideal MHD regime
\citep{JGK01, GJ02, Sisanetal04, LGJ06, SSW03}.  Liquid metals (such
as sodium, gallium, and mercury) are often characterized by rather low
magnetic Prandtl numbers (${\rm Pm}\simeq 10^{-5}$--$10^{-7}$).
Although the regime of Reynolds numbers involved is still orders of
magnitude smaller than any astrophysical system with similar magnetic
Prandtl numbers, MRI experiments offer one of the few prospects of
studying anything close to MHD astrophysical processes in the
laboratory.

A number of analyses addressing some aspects of the impact of
viscosity and resistivity on the MRI in various dissipative limits
appear scattered throughout the literature on theoretical, numerical,
and experimental MRI.  More recently, \citet{LB07} have found
particular solutions of the viscous, resistive MHD equations
(including even a cooling term) in the shearing box approximation.
However, we are unaware of any comprehensive, systematic study
addressing how the MRI behaves in viscous, resistive, differentially
rotating magnetized plasmas for arbitrary combinations of the Reynolds
and magnetic Reynolds numbers.  The aim of this work is to carry out
this analysis in detail.

The rest of paper is organized as follows.  In
\S~\ref{sec:assumptions}, we state our assumptions. In
\S~\ref{sec:solution}, we solve the eigenvalue problem defined by the
MRI for arbitrary Reynolds and magnetic Reynolds numbers.  We provide
closed analytical expressions for the eigenfrequencies and the
associated eigenvectors. In \S~\ref{sec:mri_modes}, we address the
unexplored physical structure of MRI modes for finite Reynolds and
magnetic Reynolds numbers and derive simple analytical expressions
that describe these modes in various asymptotic regimes.  In
\S~\ref{sec:stresses_energies}, we calculate the correlations between
magnetic and velocity MRI-driven perturbations that are related to
angular momentum transport and energy densities. We find that some key
results previously shown to hold in the ideal MHD limit \citep{PCP06}
are also valid in the non-ideal regime. In particular, we show that
even though the effectiveness with which the MRI disrupts the laminar
flow depends on the Reynolds and magnetic Reynolds numbers, the
instability always transports angular momentum outwards.  We also find
that magnetic perturbations dominate both the energetics and the
transport of angular momentum for any combination of the Reynolds and
magnetic Reynolds numbers. In \S~\ref{sec:discussion} we summarize our
findings and discuss the implications of our study.

\section{Assumptions}
\label{sec:assumptions}

Let us consider a cylindrical, incompressible background characterized
by an angular velocity profile $\bb{\Omega}=\Omega(r)\check{\bb{z}}$,
threaded by a vertical magnetic field $\bb{\bar{B}} = \bar{B}_z
\check{\bb{z}}$.  We work in the shearing box approximation, which
consist of a first order expansion in the variable $r-r_0$ of all the
quantities characterizing the flow at the fiducial radius $r_0$.  The
goal of this expansion is to retain the most relevant terms governing
the dynamics of the MHD fluid in a locally-Cartesian coordinate system
co-orbiting and corrotating with the background flow with local
(Eulerian) velocity $\bb{v} = r_0\,\Omega_0 \check{\bb{\phi}}$. (For a
more detailed discussion on this expansion see \citealt{GX94} and
references therein.)

The equations governing the dynamics of an incompressible MHD fluid
with constant kinematic viscosity $\nu$ and resistivity $\eta$ in the
shearing box limit are given by
\begin{eqnarray}
\label{eq:euler}
\D{t}{\bb{v}} + \left(\bb{v}\bcdot\del\right)\bb{v} & = &
- 2 \bb{\Omega}_0 \btimes \bb{v} \, + 
\, q \Omega^2_0\del(r-r_0)^2 \nonumber \\
&-& \frac{1}{\rho}\del\left(P + \frac{\bb{B}^2}{8\pi}\right)  +
\frac{(\bb{B}\bcdot\del)\bb{B}}{4\pi\rho} + \nu \del^2{\bb{v}} \,,
\nonumber \\ \\
\label{eq:induction}
\D{t}{\bb{B}} + \left( \bb{v} \bcdot \del \right)\bb{B} 
& = & \left(\bb{B} \bcdot \del \right) \bb{v} + \eta \del^2{\bb{B}} \,, 
\end{eqnarray}
where $P$ is the pressure, $\rho$ is the (constant) density, the
factor
\begin{eqnarray}
q\equiv-\left.\frac{d\ln\Omega}{d\ln r}\right|_{r_0} \,,
\end{eqnarray}
parametrizes the magnitude of the local shear, and we have defined the
(locally-Cartesian) differential operator
\begin{eqnarray}
\del & \equiv &  
\check{\bb{r}} \, \frac{\partial}{\partial r}  + 
\frac{\check{\bb{\phi}}}{r_0}\,\frac{\partial}{\partial \phi} +
\check{\bb{z}} \, \frac{\partial}{\partial z} \,,
\end{eqnarray}
where $\check{\bb{r}}$, $\check{\bb{\phi}}$, and $\check{\bb{z}}$ are,
coordinate-independent, orthonormal vectors corrotating with the
background flow at $r_0$. The continuity equation reduces to 
$\del \bcdot \bb{v}=0$ and there is no need for an equation of state
since the pressure can be determined from this condition.

We focus our attention on the dynamics of perturbations that depend
only on the vertical coordinate.  Under the current set of
assumptions, these types of perturbations are known to exhibit the
fastest growth rates in the ideal MHD case \citep{BH92, BH98, PP05}.
The equations governing the dynamics of these perturbations can be
obtained by noting that the velocity and magnetic fields given by
\begin{eqnarray}
\label{eq:mean_plus_perturbations_v}
\bb{v} &=& 
\delta v_r(z) \check{\bb{r}} + [- q \Omega_0 (r-r_0)+ \delta v_\phi(z)] 
\check{\bb{\phi}} + \delta v_z(z) \check{\bb{z}}  \,, \\
\label{eq:mean_plus_perturbations_b}
\bb{B} &=& \delta B_r(z) \check{\bb{r}}  + 
\delta B_\phi(z) \check{\bb{\phi}} + 
[\bar B_z + \delta B_z(z)] \check{\bb{z}} \,,
\end{eqnarray}
where the time dependence is implicit, constitute a family of exact,
non-linear, solutions to the viscous, resistive MHD equations
(\ref{eq:euler})-(\ref{eq:induction}).  As noted in \citet{GX94}, even
in the dissipative case, the only non-linear terms, which are present
through the perturbed magnetic energy density, are irrelevant in the
case under consideration.

We can further simplify
equations~(\ref{eq:euler})~and~(\ref{eq:induction}) by removing the
background shear flow\footnote{In the shearing box approximation, the
  dependence of the background flow on the radial coordinate is
  strictly linear and therefore viscous dissipation does not affect
  its dynamics.}  $\bb{v}_{\rm shear} = - q \Omega_0 (r-r_0)
\check{\bb{\phi}}$ and by realizing that we can take $\delta
v_z(z)=\delta B_z(z)=0$ without loss of generality. We then obtain
\begin{eqnarray}
\label{eq:vx}
\frac{\partial}{\partial t} \delta v_r
&=& 2 \Omega_0 \delta v_\phi + \frac{\bar B_z}{4\pi\rho} \,
\frac{\partial}{\partial z} \delta B_r  + \nu  \frac{\partial^2}{\partial z^2} \delta v_r \,,  \\
\label{eq:vy}
\frac{\partial}{\partial t} \delta v_\phi 
&=& - (2-q)\Omega_0 \delta v_r + \frac{\bar B_z}{4\pi\rho}  \,
\frac{\partial}{\partial z} \delta B_\phi + \nu
\frac{\partial^2}{\partial z^2} \delta v_\phi \,, \\
\label{eq:bx}
\frac{\partial}{\partial t} \delta B_r &=&  \bar B_z 
\frac{\partial}{\partial z} \delta v_r + \eta  \frac{\partial^2}{\partial z^2} \delta B_r\,,  \\
\label{eq:by}
\frac{\partial}{\partial t} \delta B_\phi &=& - q \Omega_0 \delta B_r + 
\bar B_z \frac{\partial}{\partial z} \delta v_\phi + \eta  \frac{\partial^2}{\partial z^2} \delta B_{\phi}\,,
\end{eqnarray}
where the first term on the right hand side of equation (\ref{eq:vy})
is related to the epicyclic frequency 
\begin{eqnarray}
\kappa\equiv\sqrt{2(2-q)}\, \Omega_0 \,,
\end{eqnarray}
at which the flow variables oscillate in a perturbed hydrodynamic
disk. For Keplerian rotation the parameter is $q=3/2$ and thus the
epicyclic frequency is $\kappa=\Omega_0$.

It is convenient to define the new variables $\delta b_i \equiv \delta
B_i/\sqrt{4\pi\rho}$ for $i=r,\phi$, and introduce dimensionless
quantities by considering the characteristic time- and length-scales
set by $1/\Omega_0$ and $\bar{B}_z/(\sqrt{4\pi\rho}\,\Omega_0)$.  The
equations satisfied by the dimensionless perturbations, $\delta
\tilde{v}_i$, $\delta \tilde{b}_i$, are then given by
\begin{eqnarray}
\label{eq:vx_nodim}
\partial_{\tilde t} \delta \tilde{v}_r &=& 2 \delta \tilde{v}_\phi + 
\partial_{\tilde z} \delta \tilde{b}_r + \tilde{\nu} \partial_{\tilde z}^2 \delta \tilde{v}_r\,,  \\
\label{eq:vy_nodim}
\partial_{\tilde t} \delta \tilde{v}_\phi &=& - (2-q) \delta \tilde{v}_r + 
\partial_{\tilde z} \delta \tilde{b}_\phi  + \tilde{\nu} \partial_{\tilde z}^2 \delta \tilde{v}_\phi \,, \\
\label{eq:bx_nodim}
\partial_{\tilde t} \delta \tilde{b}_r &=&  
\partial_{\tilde z} \delta \tilde{v}_r  + \tilde{\eta} \partial_{\tilde z}^2 \delta \tilde{b}_r\,,  \\
\label{eq:by_nodim}
\partial_{\tilde t} \delta \tilde{b}_\phi &=& - q \delta \tilde{b}_r + 
\partial_{\tilde z} \delta \tilde{v}_\phi  + \tilde{\eta} \partial_{\tilde z}^2 \delta \tilde{b}_\phi\,,
\end{eqnarray}
where $\tilde t$ and $\tilde z$ denote the dimensionless time and
vertical coordinate, respectively. 

The dynamics of ideal MRI modes, with $\nu=\eta=0$, is completely
determined by the dimensionless shear $q$ \citep{PCP06}. The effects
of viscous and resistive dissipation introduce two new dimensionless
quantities that alter the characteristics and evolution of the MRI.
With our choice of characteristic scales, it is natural to define the
Reynolds and magnetic Reynolds numbers characterizing the MHD flow as
\footnote{In a compressible fluid, the sound speed, $c_{\rm s}$,
  provides another natural characteristic speed to define the Reynolds
  and magnetic Reynolds numbers.  These definitions, e.g., ${\rm Re}'$
  and ${\rm Rm}'$, are related to those provided in equations
  (\ref{eq:re_number}) and (\ref{eq:rm_number}) via the plasma beta
  parameter $\beta=(2/\Gamma) (c_{\rm s}/\bar{v}_{{\rm A}z})^2$, with
  $\bar{v}_{{\rm A}z}= \bar{B}_z/\sqrt{4\pi\rho}$ the Alfv\'en speed
  in the $z$ direction, simply by ${\rm Re}' = (\Gamma\beta/2) {\rm
    Re}$ and ${\rm Rm}' = (\Gamma\beta/2) {\rm Rm}$ for a polytropic
  equation of state $P=K \rho^\Gamma$, with $K$ and $\Gamma$
  constants.}
\begin{eqnarray}
\label{eq:re_number}
{\rm Re} &\equiv& \frac{v_{{\rm A}z}^2}{\nu\Omega_0} = \frac{1}{\tilde\nu} \,, \\
\label{eq:rm_number}
{\rm Rm} &\equiv& \frac{v_{{\rm A}z}^2}{\eta\Omega_0} = \frac{1}{\tilde\eta} \,,
\end{eqnarray}
with associated magnetic Prandtl number
\begin{equation}
\label{eq:pm_number}
{\rm Pm} \equiv \frac{{\rm Rm}}{{\rm Re}}=\frac{\tilde\nu}{\tilde\eta} = \frac{\nu}{\eta} \,.
\end{equation}

In order to simplify the notation, we drop hereafter the tilde
denoting the dimensionless quantities.  In the rest of the paper, all
the variables are to be regarded as dimensionless, unless otherwise
specified.

\section{The Eigenvalue Problem for the Non-ideal  MRI\,:\\
A Formal Analytical Solution}
\label{sec:solution}

In this section we provide a complete analytical solution to the set
of equations (\ref{eq:vx_nodim})--(\ref{eq:by_nodim}) as a function of
the shear parameter $q$, (or, equivalently, the epicyclic frequency,
$\kappa$) for any set of values $(\nu,\eta)$ defining the viscosity and
resistivity.

It is convenient to work in Fourier space, as this provides the
advantage of obtaining explicitly the basis of modes that is needed to
construct the most general solution satisfying equations
(\ref{eq:vx_nodim})--(\ref{eq:by_nodim}). Taking the Fourier transform
of this set with respect to the $z$-coordinate, we obtain the matrix
equation
\begin{equation}
\partial_t \hat{\bb{\delta}}(k_n, t) = L  \, \hat{\bb{\delta}}(k_n,t) \,,
\label{eq:matrix_form} 
\end{equation}
where the vector $\hat{\bb{\delta}}(k_n, t)$ stands for 
\begin{equation}
\hat{\bb{\delta}}(k_n, t) = 
  \left[\begin{array}{c}
    \hat{\delta v_r}(k_n,t) \\ \hat{\delta v_\phi}(k_n,t) \\ 
    \hat{\delta b_r}(k_n,t) \\ \hat{\delta b_\phi}(k_n,t)
  \end{array}\right] \,,
\end{equation}
and $L$ represents the matrix
\begin{equation}
\label{eq:matrix_L}
L = 
\left[\begin{array}{cccc}
    -\nu k_n^2 & 2  & i k_n  & 0 \\
    -(2-q)  & -\nu k_n^2 & 0 & i k_n  \\
    i k_n  & 0 & -\eta k_n^2 & 0 \\
    0 & i k_n  & -q  & -\eta k_n^2
  \end{array}\right] \,. 
\end{equation}
The functions denoted by $\hat f(k_n, t)$ correspond to the Fourier
transform of the real functions, $f(z,t)$, and are defined via
\begin{equation}
\label{eq:ft_discrete}
\hat f(k_n,t) \equiv \frac{1}{2H}\int_{-H}^{H} f(z,t) \, e^{-ik_nz}
\,dz \,,
\end{equation}
where we have assumed periodic boundary conditions at $z=\pm H$, with
$H$ being the (dimensionless) scale-height and $k_n$ the wavenumber in
the $z$-coordinate,
\begin{equation}
\label{eq:invft_discrete}
k_n \equiv \frac{n\pi}{H} \,,
\end{equation}
where $n$ is an integer number. In order to simplify the notation,
hereafter we denote these wavenumbers simply by $k$.

\begin{figure*}[t]
  \includegraphics[width=0.675\columnwidth,trim=0 0 0 0]{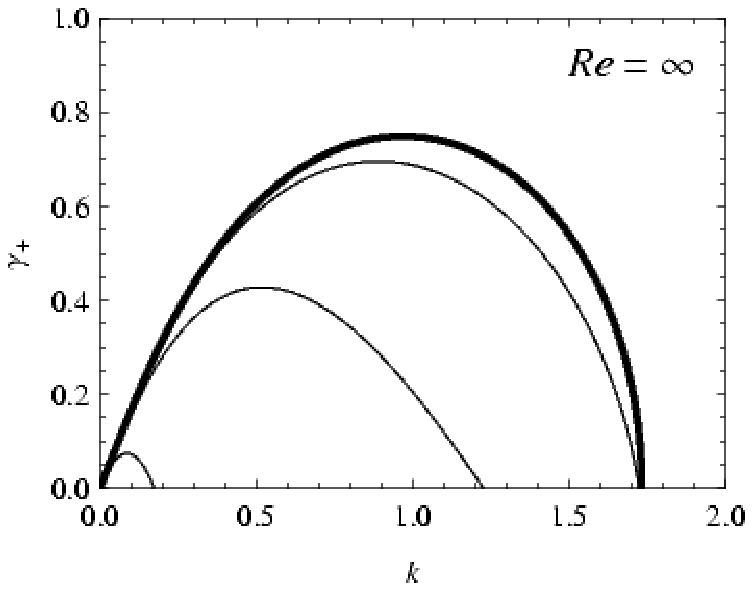}
  \includegraphics[width=0.675\columnwidth,trim=0 0 0 0]{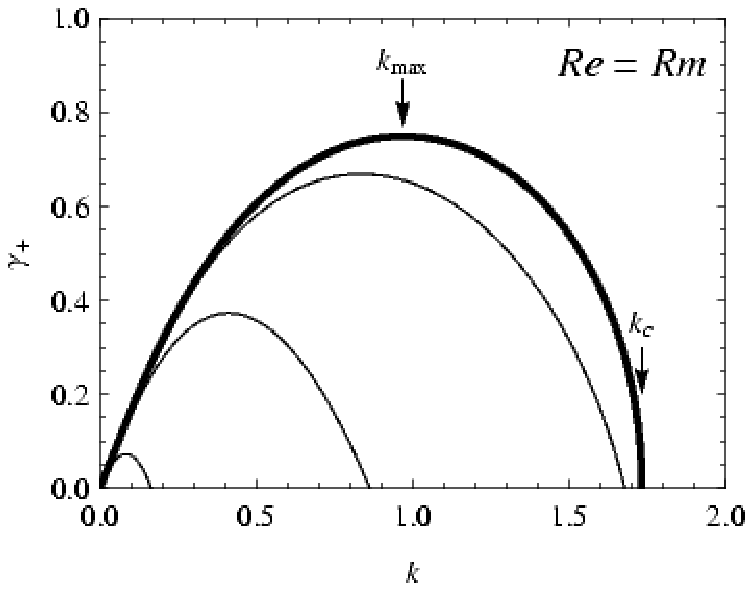}
  \includegraphics[width=0.675\columnwidth,trim=0 0 0 0]{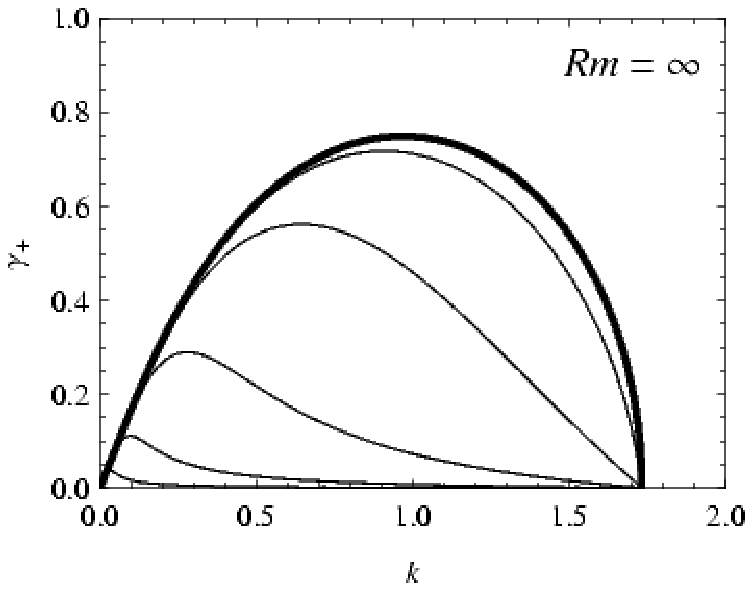}
  \caption{Growth rates $\gamma_+$,
    eq.~(\ref{eq:eigenvalues_gamma_pm}), as a function of the vertical
    wavenumber $k$ for different combinations of Reynolds and magnetic
    Reynolds numbers for Keplerian rotation.  In all three panels, the
    thick solid line corresponds to the ideal MHD limit, i.e., ${\rm
      Re}, {\rm Rm}\rightarrow\infty$.  For any combination of the
    Reynolds and magnetic Reynolds numbers, the growth rate has a well
    defined, single maximum $\gamma_{\rm max}$ that corresponds to the
    most unstable mode $k_{\rm max}$. The range of unstable modes,
    $0<k<k_{\rm c}$, is always finite, the critical wavenumber $k_{\rm
      c}$ satisfies eq.~(\ref{eq:dispersion_relation_nu_eta}) when
    $\sigma\equiv 0$, see \S~\ref{sub:modes_marginal_fastest}.
    \emph{Left}: Growth rate $\gamma_+$ for different values of the
    magnetic Reynolds number in the inviscid limit, i.e., ${\rm
      Re}\rightarrow \infty$.  The thin solid lines, in decreasing
    order, correspond to ${\rm Rm}=10,1,0.1$.  \emph{Middle}: Growth
    rate $\gamma_+$ for magnetic Prandtl number ${\rm Pm}={\rm
      Rm}/{\rm Re}=1$.  The thin solid lines, in decreasing order,
    correspond to ${\rm Re}={\rm Rm}=10,1,0.1$.  In all of the cases
    shown in the left and middle panels, the critical wavenumber,
    $k_{\rm c}$, below which unstable modes can exist decreases with
    increasing resistivity, see \S~\ref{sub:modes_re_gg_rm} and
    \S~\ref{sub:modes_re_eq_rm_ll_1} for analytic expressions of these
    marginally stable modes.  \emph{Right}: Growth rate $\gamma_+$ for
    different values of the Reynolds number in the ideal conductor
    limit, i.e., ${\rm Rm}\rightarrow \infty$. The various curves, in
    decreasing order, correspond to ${\rm Re}=10,1,\ldots,10^{-3}$.
    In this case, the range of unstable modes is insensitive to the
    Reynolds number, all the modes with wavenumbers shorter than
    $k_{\rm c}=\sqrt{2q}$ are unstable, see \S~\ref{sub:modes_ideal}.
    It is evident that the growth rates and the characteristic scales,
    both $k_{\rm max}$ and $k_{\rm c}$, are more sensitive to changes
    in the resistivity than to changes in the viscosity.  The
    simultaneous analysis of all three panels leads to the conclusion
    that viscous, resistive modes with magnetic Prandtl number equal
    to unity resemble more closely inviscid, resistive modes rather
    than viscous, conductive ones, see
    \S~\ref{sub:modes_re_eq_rm_ll_1} for the explanation of this
    behavior.}
  \label{fig:growths}
\end{figure*}

In order to solve the matrix equation (\ref{eq:matrix_form}), it is
convenient to find the eigenvector basis, $\{\mathbf{e}_j\}$ with
$j=1,2,3,4$, in which $L$ is diagonal. This basis exists for all
values of the wavenumber $k$ (i.e., the rank of the matrix $L$ is
equal to 4, the dimension of the complex space) with the possible
exception of a finite number of values of $k$.  In this basis, the
action of $L$ over the set $\{\mathbf{e}_j\}$ is equivalent to a
scalar multiplication, i.e.,
\begin{equation}
L_{\rm diag} \, \mathbf{e}_{j} = \sigma_j \,\mathbf{e}_{j}  \quad
\textrm{for} \quad j=1,2,3,4 \,,
\end{equation}
where $\{\sigma_j\}$ are complex scalars.

\subsection{Eigenvalues}

In the eigenvector basis, the matrix $L$ has a diagonal
representation $L_{\rm diag}$ = diag$(\sigma_1, \sigma_2, \sigma_3,
\sigma_4)$. The eigenvalues $\{\sigma_j\}$, with $j=1,2,3,4$, are the
roots of the characteristic polynomial associated with $L$, i.e., the
dispersion relation associated with the non-ideal MRI, which can be
written in compact form as
\begin{equation}
\label{eq:dispersion_relation_nu_eta}
  (k^2 + \sigma_\nu \sigma_\eta)^2 + \kappa^2 (k^2 + \sigma_\eta^2) -
  4 k^2 = 0 \,,
\end{equation}
where we have defined the quantities
\begin{eqnarray}
\label{eq:sigma_nu}
  \sigma_\nu &\equiv& \sigma + \nu k^2 \,, \\
\label{eq:sigma_eta}
  \sigma_\eta &\equiv& \sigma +  \eta k^2 \,.
\end{eqnarray}
The dispersion relation (\ref{eq:dispersion_relation_nu_eta}) is a
fourth order polynomial with non-zero coefficients in $\sigma$ and
$\sigma^3$. In order to find its roots it is convenient to take this
polynomial to its depressed form. This can be achieved by defining the
new variables $\sigma_\mu$ and $\mu$ such that\footnote{A physical
  interpretation of the variable $\mu$ is provided in
  \S~\ref{sub:geometrical_representation}.}
\begin{eqnarray}
\label{eq:sigma_mu}
\sigma_\mu &\equiv& \frac{1}{2}(\sigma_\nu +\sigma_\eta) \,, \\
\label{eq:mu}
\mu &\equiv& \frac{1}{2}(\nu-\eta)k^2 \,.
\end{eqnarray}
The resulting polynomial can then be written as
\begin{eqnarray}
\label{eq:dispersion_relation_mu}
\sigma_\mu^4 + \alpha \sigma_\mu^2 +\beta \sigma_\mu + \lambda  =  0 \,,
\end{eqnarray}
where the coefficients $\alpha$, $\beta$, and $\lambda$ are given by
\begin{eqnarray}
\alpha &\equiv& 2(k^2-\mu^2) + \kappa^2 \,, \\
\beta  &\equiv& -2\mu\kappa^2 \,, \\
\lambda &\equiv& (k^2-\mu^2)^2 + \kappa^2(k^2 + \mu^2) - 4k^2 \,.
\end{eqnarray}

\begin{figure*}[t]
  \includegraphics[width=\columnwidth,trim=0 0 0 0]{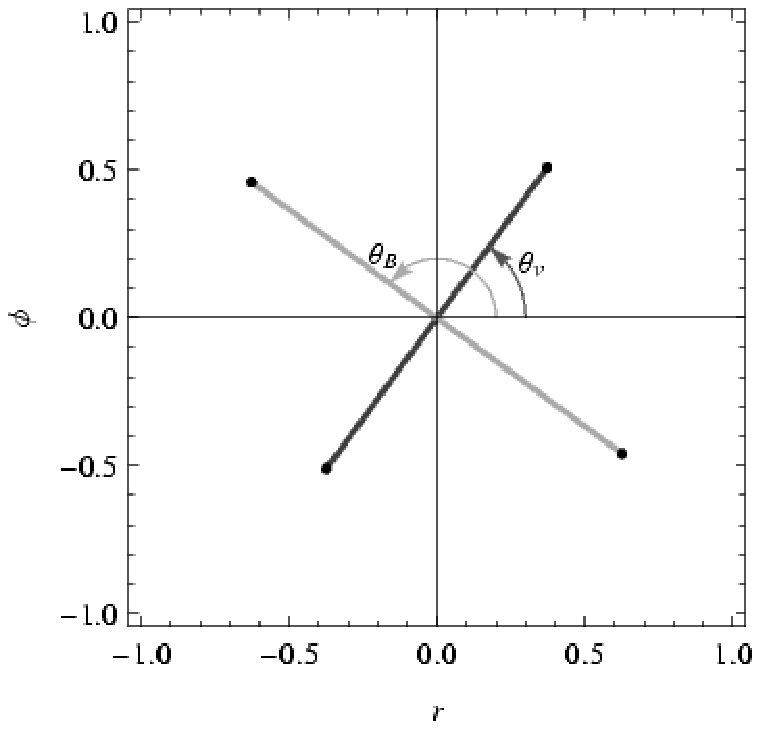}
  \includegraphics[width=\columnwidth,trim=0 0 0 0]{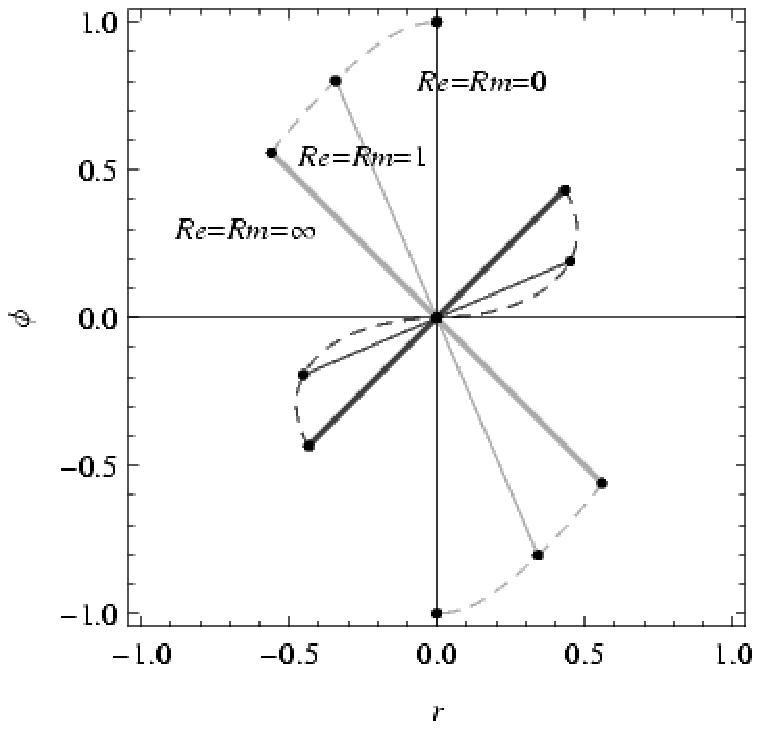}
  \caption{\emph{Left}: Geometrical representation of the velocity
    field (black) and magnetic field (gray) perturbations for viscous,
    resistive MRI modes.  Note that this is a projection of a single
    mode, which is inherently three-dimensional, onto the disk
    mid-plane $(r,\phi,z=0)$. The velocity and magnetic field
    components are always out of phase in the vertical direction $z$
    by $\pi/4$, see eq.~(\ref{eq:delta_zt_unstable}). The angles
    $\theta_{\rm v}$ and $\theta_{\rm b}$, defined in
    eqs.~(\ref{eq:theta_v})~and~(\ref{eq:theta_b}), respectively,
    correspond to the physical angles defining the planes
    (perpendicular to the disk midplane) containing the MRI-driven
    perturbations, see
    eqs.~(\ref{eq:tan_theta_v})~and~(\ref{eq:tan_theta_b}).  The
    relative magnitude of velocity and magnetic field perturbations is
    determined by eq.~(\ref{eq:b0_v0}).  \emph{Right}: Evolution of
    the geometrical representation of the fastest-growing, non-ideal
    MRI mode, with associated wavenumber $k_{\rm max}$, with magnetic
    Prandtl number equal to unity, as a function of the
    Reynolds/magnetic Reynolds number. When the Reynolds/magnetic
    Reynolds number varies according to ${\rm Re}={\rm
      Rm}:\infty\rightarrow 0$, the angles evolve according to
    $\theta_{\rm v}:\pi/4\rightarrow 0$ and $\theta_{\rm
      b}:3\pi/4\rightarrow \pi/2$ and the relative amplitude of the
    perturbations evolves according to $b_0/v_0:5/3\rightarrow
    \infty$.  Note that the velocity and magnetic field perturbations
    are always orthogonal for ${\rm Pm}=1$, see
    eq.~(\ref{eq:theta_bv}).}
  \label{fig:geom_def}
\end{figure*}

The solutions to equation (\ref{eq:dispersion_relation_mu}) are
\begin{eqnarray}
  \sigma_\mu = \pm_a (-\Lambda \mp_b \sqrt{\Delta})^{1/2} \pm_b \frac{\beta}{4\sqrt{\Delta}} \,,
\end{eqnarray}
where the subscripts $a$ and $b$ in the ``$+$'' and ``$-$'' signs
label the four possible combinations of signs and we have defined the
quantities\footnote{Defining the quantities $\Lambda$ and $\Delta$ in
  this way allows us to show explicitly that in the limit $\nu, \eta
  \rightarrow 0$ the solutions to equation
  (\ref{eq:dispersion_relation_nu_eta}) converge smoothly to the
  solutions found in the ideal MHD case (see Appendix
  \ref{sec:appendix}).}
\begin{eqnarray}
  \Lambda &=& \frac{3\alpha}{4} + \frac{y}{2} \,, \\
  \Delta &=& (y + \alpha)^2 - \lambda \,,
\end{eqnarray}
and $y$ is any of the solutions to the cubic equation
\begin{equation}
  y^3 + \frac{5\alpha}{2} y^2 + (2\alpha^2 - \lambda) y +
  \left(\frac{\alpha^3}{2} - \frac{\alpha\lambda}{2} -
  \frac{\beta^2}{8}\right) = 0 \,,
\end{equation}
which has closed analytic solutions
\begin{equation}
  y = -\frac{5}{6} \alpha + \frac{1}{3}\frac{P}{U} -U \,,
\end{equation}
with 
\begin{eqnarray}
  P &=& -\frac{\alpha^2}{12}-\lambda \,, \\
  Q &=& -\frac{\alpha^3}{108}+\frac{\alpha\lambda}{3}-\frac{\beta^2}{8} \,, \\
  U &=& \left(\frac{Q}{2}\pm \sqrt{\frac{Q^2}{4}+\frac{P^3}{27}}\right)^{1/3} \,.
\end{eqnarray}
Note that the choice of either sign in $U$ is immaterial.

It is now trivial to write the solutions of the dispersion relation
(\ref{eq:dispersion_relation_nu_eta}), $\sigma$, in terms of the
variable $\sigma_\mu$.  Using equations
(\ref{eq:sigma_nu})--(\ref{eq:sigma_mu}) we obtain
\begin{eqnarray}
\sigma =  \sigma_\mu - \frac{1}{2}(\nu+\eta)k^2 \,.
\end{eqnarray}
The eigenfrequencies of the viscous, resistive MRI modes 
are then given by
\begin{eqnarray}
\label{eq:eigenvalues_sigma}
  \sigma &=&  \pm_a (-\Lambda \mp_b \sqrt{\Delta})^{1/2} \nonumber \\
  &-&\frac{\nu}{2}  k^2 \left(1 \pm_b \frac{\kappa^2}{2\sqrt{\Delta}} \right)
  -\frac{\eta}{2} k^2 \left(1 \mp_b \frac{\kappa^2}{2\sqrt{\Delta}}
  \right) \,.
\end{eqnarray}

\subsection{Four classes of solutions}
\label{sub:classes_of_solutions}

All of the quantities $\Lambda$, $\Delta$, and $y$, depend on the
viscosity $\nu$ and the resistivity $\eta$ only through $\mu^2 \propto
(\nu-\eta)^2$. This has a series of important implications, in
particular, there is always a range of wavenumbers for which the
discriminant in equation (\ref{eq:eigenvalues_sigma}) is positive,
i.e., $\sqrt{\Delta} - \Lambda > 0$.  It can also be seen that the
last two terms between parentheses in equation
(\ref{eq:eigenvalues_sigma}) are always positive, i.e., $\sqrt{\Delta}
\ge \kappa^2/2$, and thus they always produce damping.  Because of
this, we can classify the modes in four types: two (damped) growing
and decaying ``unstable'' modes with eigenvalues
\begin{eqnarray}
\label{eq:eigenvalues_gamma_pm}
  \gamma_{\pm} &=&  \pm (\sqrt{\Delta}-\Lambda)^{1/2} \nonumber \\
  &-&\frac{\nu}{2}  k^2 \left(1 - \frac{\kappa^2}{2\sqrt{\Delta}} \right)
  -\frac{\eta}{2} k^2 \left(1 + \frac{\kappa^2}{2\sqrt{\Delta}}
  \right) \,,
\end{eqnarray}
and two (damped) ``oscillatory'' modes with eigenvalues
\begin{eqnarray}
\label{eq:eigenvalues_omega_pm}
 i \omega_{\pm} &=&  \pm i (\sqrt{\Delta}+\Lambda)^{1/2} \nonumber \\
  &-&\frac{\nu}{2}  k^2 \left(1 + \frac{\kappa^2}{2\sqrt{\Delta}} \right)
  -\frac{\eta}{2} k^2 \left(1 - \frac{\kappa^2}{2\sqrt{\Delta}}
  \right) \,.
\end{eqnarray}
We arbitrarily label these eigenvalues as
\begin{eqnarray}
\label{eq:eigenvalues_all}
\sigma_{1} \equiv \gamma_{+} \,,   \quad
\sigma_{2} \equiv \gamma_{-} \,,   \quad
\sigma_{3} \equiv i \omega_{+} \,, \quad 
\sigma_{4} \equiv i \omega_{-} \,.  
\end{eqnarray}

\begin{figure*}[t]
  \includegraphics[width=0.675\columnwidth,trim=0 0 0 0]{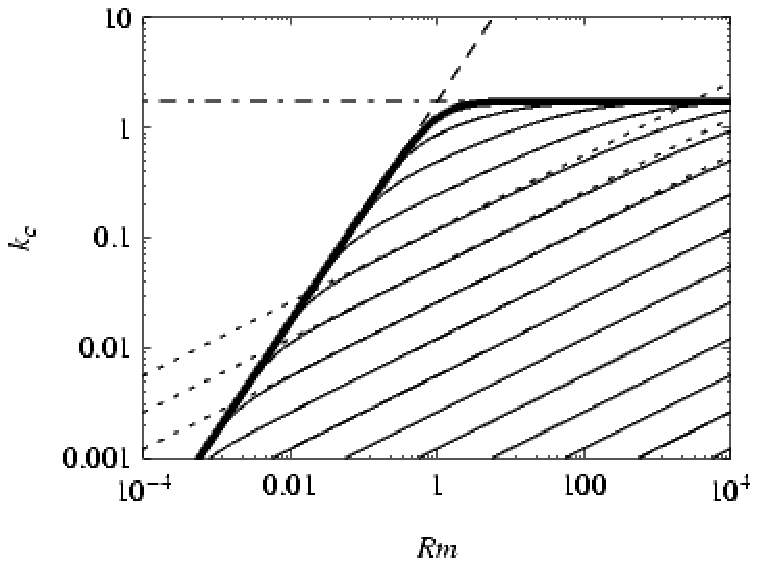}
  \includegraphics[width=0.675\columnwidth,trim=0 0 0 0]{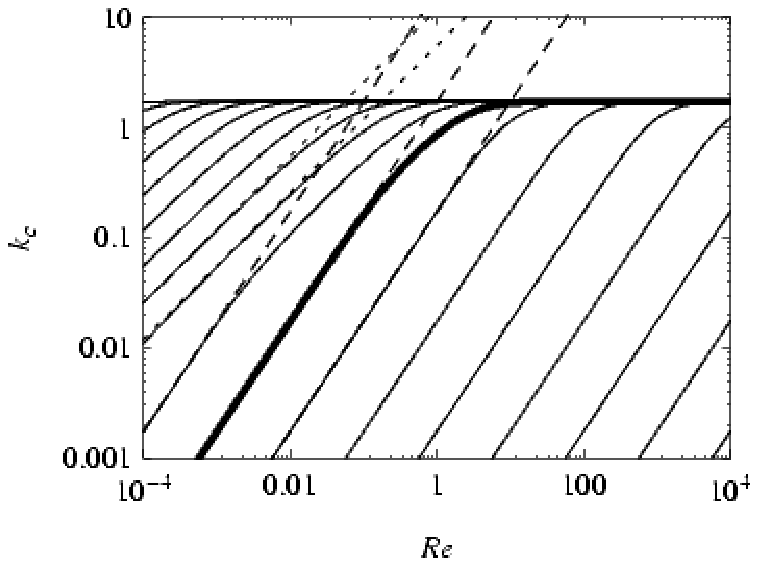}
  \includegraphics[width=0.675\columnwidth,trim=0 0 0 0]{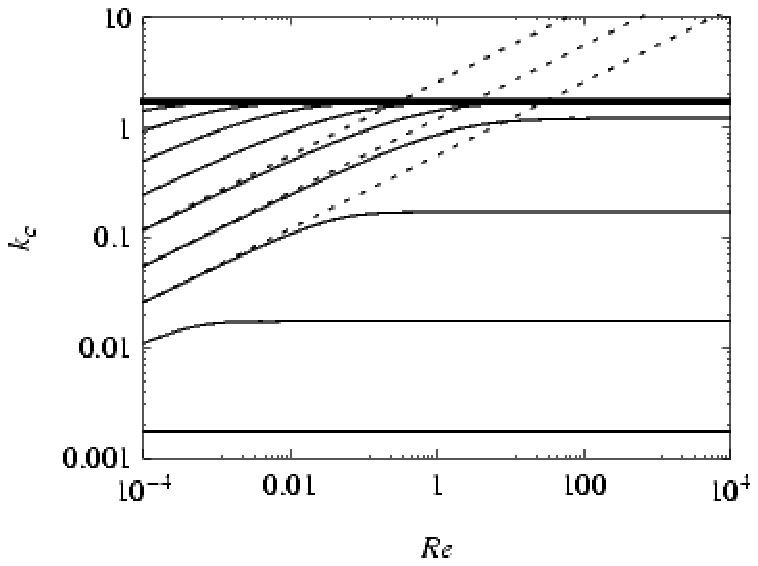}
  \caption{Critical wavenumber $k_{\rm c}$, see eq.~(\ref{eq:k_c}),
    corresponding to the marginally stable MRI mode for Keplerian
    rotation in different dissipative regimes.  The horizontal lines
    at $k_{\rm c}=\sqrt{3}$ represent the ideal MHD limit,
    eq.~(\ref{eq:k_c_ideal}).  \emph{Left}: Critical wavenumber
    $k_{\rm c}$ as a function of the magnetic Reynolds number for
    different values of the Reynolds number.  The thick solid line
    denotes the inviscid limit, i.e., ${\rm Re}\rightarrow \infty$.
    Note that eq.~(\ref{eq:k_c_lim_reggrm}) describes this curve
    exactly. The thin solid lines, in decreasing order, correspond to
    ${\rm Re} = 1, 0.1, \ldots$. For small magnetic Reynolds number
    $k_{\rm c} \propto {\rm Rm}$, see eq.~(\ref{eq:k_c_lim_reggrm}).
    For finite Reynolds numbers, such that ${\rm Re\,Rm} \lesssim 1$
    there is a transition between the regimes ${\rm Rm}\ll 1$ and
    ${\rm Rm}\gg 1$ such that $k_{\rm c} \propto {\rm Rm}^{1/3}$, see
    eq.~(\ref{eq:k_c_lim_rellrm}).  \emph{Middle}: Critical wavenumber
    $k_{\rm c}$ as a function of the Reynolds number for different
    magnetic Prandtl numbers.  ${\rm Pm}$ increases/decreases by an
    order of magnitude for each curve to the left/right of the thick
    solid line denoting the ${\rm Pm}=1$ case.  The dashed lines
    $k_{\rm c} \propto {\rm Re\,Pm} = {\rm Rm}$ are calculated
    according to eq.~(\ref{eq:k_c_lim_reggrm}), which gives the
    correct result even for ${\rm Pm} > 1$, provided that the Reynolds
    number is sufficiently small.  The dotted lines $k_{\rm c} \propto
    ({\rm Re\,Rm})^{1/3}$ are calculated according to eq.~
    (\ref{eq:k_c_lim_rellrm}).  \emph{Right}: Critical wavenumber
    $k_{\rm c}$ as a function of the Reynolds number for different
    values of the magnetic Reynolds number.  The thick solid line
    corresponds to the ideal conductor limit, i.e., ${\rm
      Rm}\rightarrow \infty$.  The thin solid lines, in decreasing
    order, correspond to ${\rm Re} =10^5, 10^4, \ldots, 10^{-3}$.  For
    small Reynolds numbers $k_{\rm c} \propto {\rm Re}^{1/3}$, see
    eq.~(\ref{eq:k_c_lim_rellrm}), while for Reynolds numbers larger
    than a few, the critical wavenumber is independent of ${\rm Re}$
    regardless of the value of ${\rm Rm}$.}
  \label{fig:k_c}
\end{figure*}

Figure~\ref{fig:growths} shows the growth rate $\gamma_+$ as a
function of the vertical wavenumber $k$ for different combinations of
the Reynolds and magnetic Reynolds numbers for Keplerian rotation.
These growth rates are more sensitive to changes in the resistivity
than to changes in the viscosity.  A qualitative understanding of this
behavior can be obtained by realizing that viscosity tends to quench
the instability, without altering the large scale magnetic field.
Thus, as long as the resistivity is negligible, the range of unstable
lenghtscales are the same in both ideal and viscous, perfectly
conducting fluids.  On the other hand, resistivity tends to destroy
the magnetic field at small scales having a stronger impact on the
stability of the perturbations at these scales.

Mathematically, the asymmetric response of the growth rate to changes
in the viscosity $\nu$ or the resistivity $\eta$ originates in the
different functional form of the terms that contribute to produce
damping, i.e., $(1-\kappa^2/2\sqrt{\Delta})$ and
$(1+\kappa^2/2\sqrt{\Delta})$, in the exponential growth characterized
by $\gamma_+$ in equation (\ref{eq:eigenvalues_gamma_pm}). If the
oscillatory modes, $\omega_{\pm}$ in equation
(\ref{eq:eigenvalues_omega_pm}), are considered instead, the roles of
the plus and minus signs in these terms are interchanged. From this
analysis we can infer that the ``oscillatory'' mode is affected
(damped) more strongly by viscosity than by resistivity.

The simultaneous analysis of the various panels in
Figure~\ref{fig:growths} leads to the conclusion that viscous,
resistive unstable modes with magnetic Prandtl number equal to unity
resemble more closely inviscid, resistive modes rather than viscous,
conductive ones.  In \S~\ref{sec:mri_modes} we provide analytical
expressions to support this conclusion.

\subsection{Normalized Eigenvectors: Geometrical Representation}
\label{sub:geometrical_representation}

The set of normalized eigenvectors, $\{\mathbf{e}_{\sigma_j}\}$,
associated with the eigenvalues (\ref{eq:eigenvalues_all}) are given by
\begin{eqnarray}
\label{eq:eigenvectors_initial}
\mathbf{e}_{\sigma_j} &\equiv& \frac{\mathbf{e}_{j}}{\|\mathbf{e}_{j}\|}
\quad \textrm{for} \quad j=1,2,3,4 \,,
\end{eqnarray}
where
\begin{equation}
\label{eq:e_sigma_j}
\mathbf{e}_{j}(k) =
    \left[\begin{array}{c}
      \sigma_{\eta j} \\
      (k^2  + \sigma_{\nu j} \sigma_{\eta j})/2 \\
      i k  \\
      - i k  [2 \sigma_{\eta j} + q (\nu - \eta)k^2
      ]/(k^2  + \sigma_{\nu j} \sigma_{\eta j})
    \end{array}\right],
\end{equation}
$\sigma_{\nu j} = \sigma_j + \nu k^2$, $\sigma_{\eta j} = \sigma_j +
\eta k^2$, and the norms are given by
\begin{equation}
  \|\mathbf{e}_{j}\| 
  \equiv \left[\sum_{l=1}^{4} 
    {\rm e}^{l}_{j} 
    {\rm e}^{l*}_{j}\right]^{1/2} \,.
\end{equation}
Here, ${\rm e}^{l}_{j}$ is the $l$-th component of the (unnormalized)
eigenvector associated with the eigenvalue $\sigma_j$.  

The set of four eigenvectors $\{\mathbf{e}_{\sigma_j}\}$, together
with the set of complex scalars $\{\sigma_j\}$ in equation
(\ref{eq:eigenvalues_all}), constitute the full solution to the
eigenvalue problem defined by the MRI for any combination of the
Reynolds and magnetic Reynolds numbers.

A geometrical representation of the eigenvectors
(\ref{eq:eigenvectors_initial}) can be brought to light by defining
the angles $\theta_{{\rm v}j}$ and $\theta_{{\rm b}j}$ according to
\begin{eqnarray}
\label{eq:theta_v}
\tan \theta_{{\rm v}j} &\equiv& \frac{e^2_j}{e^1_j} 
= \frac{k^2+ \sigma_{\nu j} \sigma_{\eta j}}{2\sigma_{\eta j}} \,, \\
\label{eq:theta_b}
\tan \theta_{{\rm b}j} &\equiv& \frac{e^4_j}{e^3_j} 
= - \frac{2\sigma_{\eta j} + q (\nu-\eta)k^2}{k^2+ \sigma_{\nu j} \sigma_{\eta j}}\,.
\end{eqnarray}
It is important to remark that each of the four eigenvectors define,
in principle, four sets of angles $\{\theta_{{\rm v}j},\theta_{{\rm
    b}j}\}$, for $j=1,2,3,4$.  We label the angles associated with the
different types of modes discussed in
\S~\ref{sub:classes_of_solutions} according to
\begin{eqnarray}
\label{eq:theta_all}
\theta_{{\rm v}1} \equiv \theta_{\rm v}^{\gamma_+},   \quad
\theta_{{\rm v}2} \equiv \theta_{\rm v}^{\gamma_-},   \quad
\theta_{{\rm v}3} \equiv \theta_{\rm v}^{\omega_+},   \quad
\theta_{{\rm v}4} \equiv \theta_{\rm v}^{\omega_-}, \,\,\,\,\,
\end{eqnarray}
with similar definitions corresponding to $\theta_{{\rm b}j}$ for
$j=1,2,3,4$.  Note that these angles are defined in \emph{spectral}
space and depend, in general, on the wavenumber $k$, the epicyclic
frequency, $\kappa$, the viscosity $\nu$, and the resistivity $\eta$.
The angles associated with the modes labeled by $\gamma_+$ and
$\gamma_-$ are always real while the ones associated with the modes
$\omega_+$ and $\omega_-$ are in general complex.  For the sake of
brevity, in what follows we will refer to the set of angles describing
unstable MRI modes $\{\theta_{\rm v}^{\gamma_+}, \theta_{\rm
  b}^{\gamma_+}\}$ simply as $\{\theta_{\rm v}, \theta_{\rm b}\}$.

\begin{figure*}[t]
  \includegraphics[width=0.675\columnwidth,trim=0 0 0 0]{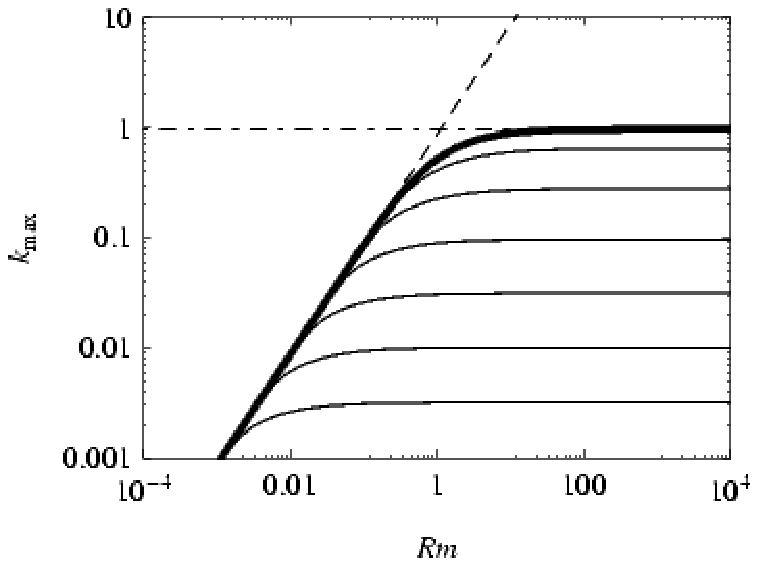}
  \includegraphics[width=0.675\columnwidth,trim=0 0 0 0]{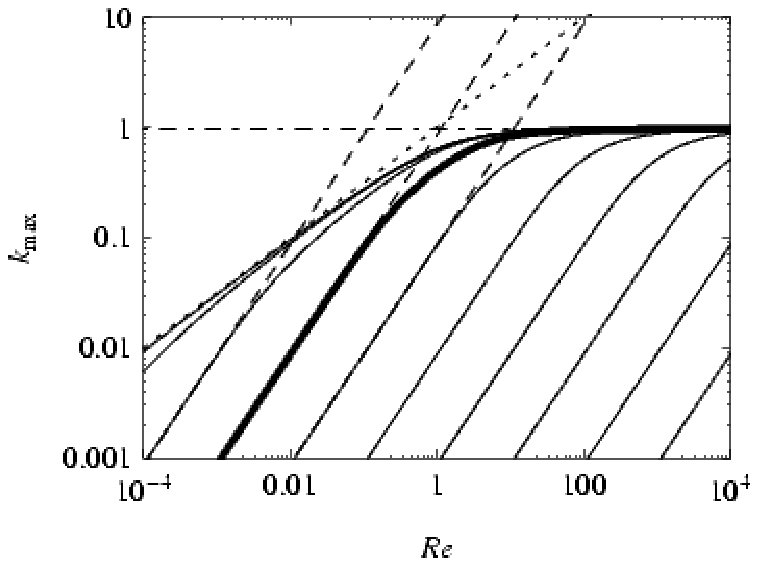}
  \includegraphics[width=0.675\columnwidth,trim=0 0 0 0]{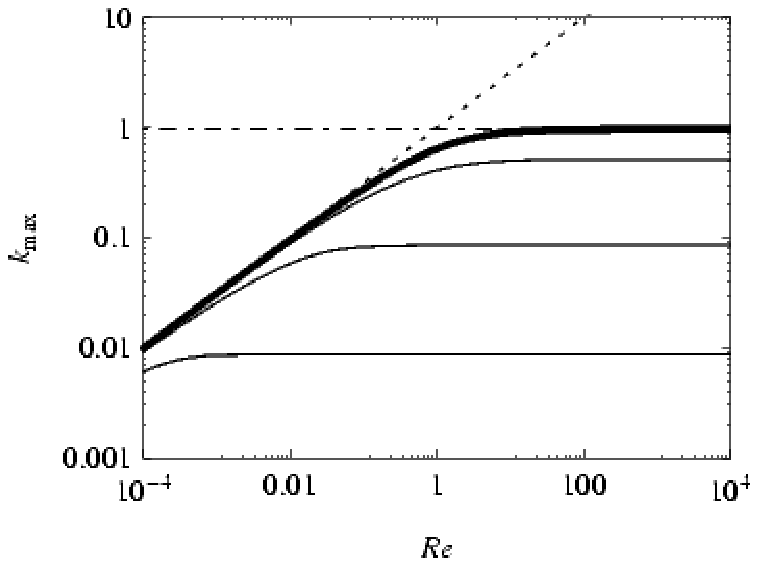}
  \caption{Wavenumber $k_{\rm max}$ corresponding to the fastest
    growing non-ideal MRI modes for Keplerian rotation in different
    dissipative regimes.  The dot-dashed horizontal lines at $k_{\rm
      max}=\sqrt{15/16}$ represent the ideal MHD limit,
    eq.~(\ref{eq:k_max_ideal}). \emph{Left}: Fastest growing mode
    $k_{\rm max}$ as a function of the magnetic Reynolds number for
    different values of the Reynolds number.  The thick solid line
    denotes the inviscid limit, i.e., ${\rm Re}\rightarrow \infty$.
    The thin solid lines, in decreasing order, correspond to ${\rm Re}
    =1, 0.1, \ldots, 10^{-5}$.  For magnetic Reynolds numbers larger
    than unity, this wavenumber is independent of ${\rm Rm}$
    regardless of the value of ${\rm Re}$.  The dashed line,
    calculated according to eq.~(\ref{eq:k_max_lim_reggrm}), provides
    the correct asymptotic limit $k_{\rm max} \propto {\rm Rm}$ for
    small magnetic Reynolds numbers.  \emph{Middle}: Fastest growing
    mode $k_{\rm max}$ as a function of the Reynolds number for
    different values of the magnetic Prandtl number.  From left to
    right, the curves correspond to ${\rm Pm}=10^3, 10^2, \ldots, 1$
    (thick solid line), $\ldots, 10^{-6}$. The dashed lines $k_{\rm
      max} \propto {\rm Re\,Pm}={\rm Rm}$ are calculated according to
    eq.~(\ref{eq:k_max_lim_reggrm}), which leads to the correct result
    even for ${\rm Pm} \gtrsim 1$ provided that the Reynolds number is
    sufficiently small.  The dotted line $k_{\rm max} \propto {\rm
      Re}^{1/2}$ results from eq.~(\ref{eq:k_max_lim_rellrm}).
    \emph{Right}: Fastest growing mode $k_{\rm max}$ as a function of
    the Reynolds number for different values of the magnetic Reynolds
    number.  The thick solid line corresponds to the ideal conductor
    limit, i.e., ${\rm Rm}\rightarrow \infty$.  The thin solid lines,
    in decreasing order, correspond to ${\rm Rm} = 1, 0.1, 0.01$.  For
    Reynolds numbers larger than unity, the growth rate is independent
    of ${\rm Re}$ regardless of the value of ${\rm Rm}$.  The dotted
    line, calculated according to eq.~(\ref{eq:k_max_lim_rellrm}),
    provides the correct asymptotic limit $k_{\rm max} \propto {\rm
      Re}^{1/2}$ for small Reynolds number.}
  \label{fig:k_max}
\end{figure*}

A normalized version of the MRI eigenvectors can be obtained by
multiplying the set of vectors in equation (\ref{eq:e_sigma_j}) by the
amplitudes
\begin{eqnarray}
  A_j \equiv\sqrt{\frac{2}{q(k^2+\sigma_{\eta j}^2)}}  \, \frac{v_0}{\sqrt{v_0^2+b_{0}^2}} \,.
\end{eqnarray}
where we have defined
\begin{eqnarray}
\label{eq:b0_v0}
b_{0}\equiv \frac{2k v_0}{k^2+\sigma_{\nu j} \sigma_{\eta j}} \, \left(1 +
  \frac{(\nu-\eta)[4\sigma_{\eta j} +
    q(\nu-\eta)k^2]k^2}{2(k^2+\sigma_{\eta j}^2)}\right)^{1/2} \,,
\nonumber \\ 
\end{eqnarray}
where, for the sake of simplicity, we have omitted the subscript $j$
on the left hand side. The expressions for the normalized eigenvectors
$\{\mathbf{e}_{\sigma_j}\}$, for $j=1,2,3,4$, are then given by
\begin{equation}
\label{eq:e_normalized}
\mathbf{e}_{\sigma_j}(k) = 
 \frac{1}{\sqrt{v_0^2+b_0^2}}
  \left[\begin{array}{r}
       v_0 \cos\theta_{{\rm v}j} \\ 
       v_0 \sin\theta_{{\rm v}j} \\ 
       i b_0 \cos\theta_{{\rm b}j} \\ 
       i b_0 \sin\theta_{{\rm b}j}
  \end{array}\right] \,.
\end{equation}

It is interesting to note that in this geometric representation the
dispersion relation (\ref{eq:dispersion_relation_nu_eta}) can be
obtained from the trigonometric identity
\begin{eqnarray}
\cos^2\theta_{{\rm v}j} + \sin^2\theta_{{\rm v}j}=1 \,,
\end{eqnarray}
where the expressions for 
\begin{eqnarray}
\cos\theta_{{\rm v}j} &=& \sqrt{\frac{2 \sigma_{\eta j}^2}{q(k^2+\sigma_{\eta j}^2)}} \,, \\
\sin\theta_{{\rm v}j} &=& \frac{k^2+\sigma_{\nu j}\sigma_{\eta j}}{\sqrt{2q(k^2+\sigma_{\eta j}^2)}} \,,
\end{eqnarray}
can be obtained from the definition of the angle $\theta_{{\rm v}j}$ in
equation (\ref{eq:theta_v}).

%\newpage 
\subsection{Temporal Evolution}

In physical space, the most general solution to the set of equations
(\ref{eq:vx_nodim})--(\ref{eq:by_nodim}), i.e.,
\begin{equation}
\bb{\delta}(z, t) = 
  \left[\begin{array}{c}
    \delta v_r(z,t) \\ \delta v_\phi(z,t) \\ 
    \delta b_r(z,t) \\ \delta b_\phi(z,t)
  \end{array}\right] \,,
\end{equation}
evolves in time according to 
\begin{equation}
\label{eq:solution_0}
{\bb \delta}(z,t) \equiv \sum_{k} \hat {\bb \delta}(k,t) \, e^{ikz}  \,,
\end{equation}
where
\begin{equation}
\label{eq:solution_1}
\hat{\bb{\delta}}(k,t) = \sum_{j=1}^{4}  a_j(k,0) \,
e^{\sigma_j t} \,\mathbf{e}_{\sigma_j}  \,,
\end{equation}
with $\{\sigma_j\}$ and $\{\mathbf{e}_{\sigma_j}\}$, for $j=1,2,3,4$,
given by equations (\ref{eq:eigenvalues_all}) and
(\ref{eq:e_normalized}). The initial conditions $\bb{a}(k,0)$ are
related to the initial spectrum of perturbations,
$\hat{\bb{\delta}}(k,0)$, via $\bb{a}(k,0) = Q^{-1} \,
\hat{\bb{\delta}}(k,0)$. Here, $Q^{-1}$ is the matrix for the change
of coordinates from the standard basis to the normalized eigenvector
basis\footnote{The eigenvectors (\ref{eq:e_normalized}) are not in
  general orthogonal, i.e., $\mathbf{e}_{\sigma_j} \bcdot \,
  \mathbf{e}_{\sigma_{j'}} \ne 0$ for $j \ne j'$.  If desired, an
  orthogonal basis can be constructed using the Gram-Schmidt
  orthogonalization procedure \cite[see, e.g.,][]{HK71}.} and can be
obtained by calculating the inverse of the matrix
\begin{equation}
Q = [
\mathbf{e}_{\sigma_1} \ \ \mathbf{e}_{\sigma_2} \ \   
\mathbf{e}_{\sigma_3} \ \ \mathbf{e}_{\sigma_4}] \,.
\end{equation}

The temporal evolution of a single MRI-unstable mode in physical space
can be obtained from a linear combination of
$\mathbf{e}_{\sigma_j}(k)$ and $\mathbf{e}_{\sigma_j}(-k)$ as defined
in equation (\ref{eq:e_normalized}). In particular, setting
$a_1(k,0)=a^*_1(-k,0)=-i/\sqrt{2}$ in equation~(\ref{eq:solution_0})
and substituting the result in equation~(\ref{eq:solution_1}) we
obtain
\begin{equation}
\label{eq:delta_zt_unstable}
\bb{\delta}(z, t) = \, 
\frac{\sqrt{2} \, e^{\gamma_+ t}}{\sqrt{v_0^2+b_0^2}} \,
  \left[\begin{array}{r}
     v_0 \cos\theta_{{\rm v}}\, \sin(kz) \\ 
     v_0 \sin\theta_{{\rm v}}\, \sin(kz) \\ 
     b_0 \cos\theta_{{\rm b}}\, \cos(kz) \\ 
     b_0 \sin\theta_{{\rm b}}\, \cos(kz)
  \end{array}\right] \,.
\end{equation}

\begin{figure*}[t]
  \includegraphics[width=0.675\columnwidth,trim=0 0 0 0]{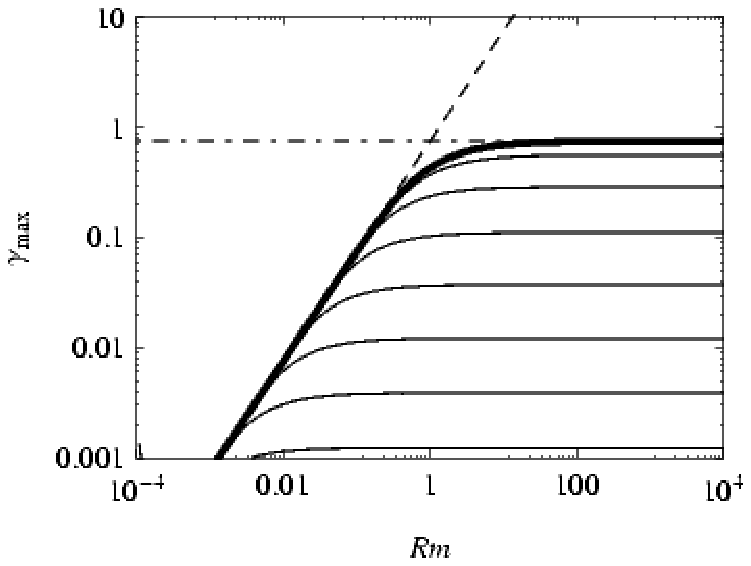}
  \includegraphics[width=0.675\columnwidth,trim=0 0 0 0]{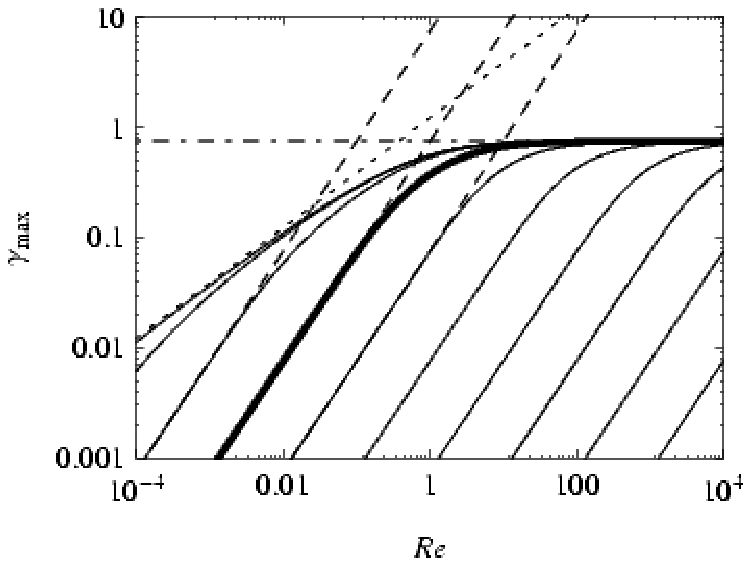}
  \includegraphics[width=0.675\columnwidth,trim=0 0 0 0]{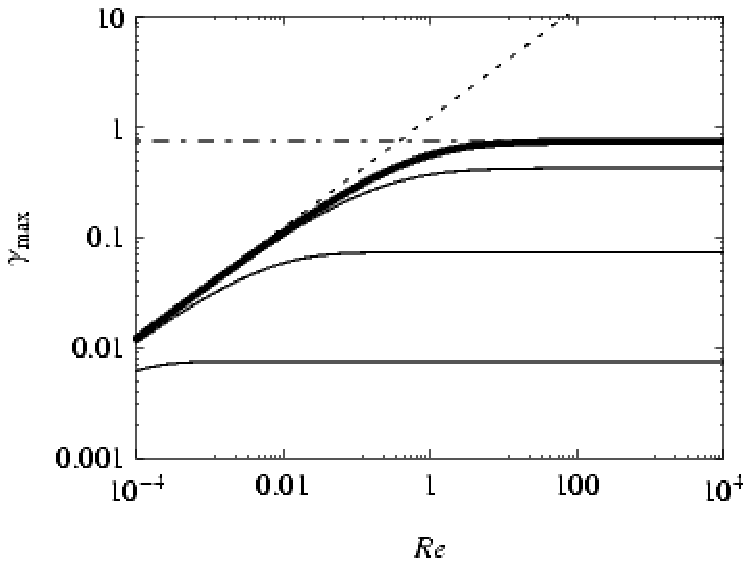}
  \caption{Maximum growth rate $\gamma_{\rm max}$ for Keplerian
    rotation in different dissipative regimes.  The dot-dashed
    horizontal lines at $\gamma_{\rm max}=3/4$ represent the ideal MHD
    limit, eq.~(\ref{eq:gamma_max_ideal}).  \emph{Left}: Maximum
    growth rate as a function of the magnetic Reynolds number for
    different values of the Reynolds number.  The thick solid line
    denotes the inviscid limit, i.e., ${\rm Re}\rightarrow \infty$.
    The thin solid lines, in decreasing order, correspond to ${\rm Re}
    =1, 0.1, \ldots, 10^{-6}$.  For magnetic Reynolds numbers larger
    than unity, the growth rate is independent of ${\rm Rm}$
    regardless of the value of ${\rm Re}$.  The dashed line,
    calculated according to eq.~(\ref{eq:gamma_max_lim_reggrm}),
    provides the correct asymptotic limit $\gamma_{\rm max} \propto
    {\rm Rm}$ for small magnetic Reynolds numbers.  \emph{Middle}:
    Maximum growth rate $\gamma_{\rm max}$ as a function of the
    Reynolds number for different values of the magnetic Prandtl
    number.  From left to right, the curves correspond to ${\rm
      Pm}=10^3, 10^2, \ldots, 1$ (thick solid line), $\ldots,
    10^{-6}$. The dashed lines $\gamma_{\rm max} \propto {\rm
      Re\,Pm}={\rm Rm}$ are calculated according to
    eq.~(\ref{eq:gamma_max_lim_reggrm}), which leads to the correct result
    even for ${\rm Pm} \gtrsim 1$ provided that the Reynolds number is
    sufficiently small.  The dotted line $\gamma_{\rm max} \propto
    {\rm Re}^{1/2}$ results from eq.~(\ref{eq:gamma_max_lim_rellrm}).
    \emph{Right}: Maximum growth rate $\gamma_{\rm max}$ as a function
    of the Reynolds number for different values of the magnetic
    Reynolds number.  The thick solid line corresponds to the ideal
    conductor limit, i.e., ${\rm Rm}\rightarrow \infty$.  The thin
    solid lines, in decreasing order, correspond to ${\rm Rm} = 1,
    0.1, 0.01$.  For Reynolds numbers larger than unity, the growth
    rate is independent of ${\rm Re}$ regardless of the value of ${\rm
      Rm}$.  The dotted line, calculated according to
    eq.~(\ref{eq:gamma_max_lim_rellrm}), provides the correct
    asymptotic limit $\gamma_{\rm max} \propto {\rm Re}^{1/2}$ for
    small Reynolds number.}
  \label{fig:gamma_max}
\end{figure*}

These solutions are of particular importance for the linear late-time
evolution of MRI modes.  Note that any reasonable spectrum of initial
perturbations of the type used in numerical simulations of shearing
boxes will have a non-zero component along the unstable eigenvector
$e_{\sigma_1}$.  If the value of the magnetic field is such that the
MRI can be excited for given values of the viscosity and resistivity
then the exponentially growing perturbations in physical space will
evolve towards a mode of the form (\ref{eq:delta_zt_unstable})
dominated by the lengthscale $k=k_{\rm max}$ for which the growth rate
reaches its maximum value $\gamma_{\rm max}$.

Note that if a perturbation in physical space is composed by a single
mode of the type described in \S~\ref{sub:classes_of_solutions}, no
matter which class, then the angles defined in equations
(\ref{eq:theta_v}) and (\ref{eq:theta_b}) are constant in time and are
identical to the \emph{physical} angles between the planes containing
magnetic and velocity perturbations in physical space, see
Figure~\ref{fig:geom_def}, with
\begin{eqnarray}
\label{eq:tan_theta_v}
\tan\theta_{{\rm v}j} &=& \frac{\delta v_\phi(z,t)}{\delta v_r(z,t)}={\rm const.} \,, \\
\label{eq:tan_theta_b}
\tan\theta_{{\rm b}j} &=& \frac{\delta b_\phi(z,t)}{\delta b_r(z,t)}={\rm const.}\,.
\end{eqnarray}

Finally, defining the angle $\theta_{{\rm bv}j}$ such that
\begin{eqnarray}
\theta_{{\rm bv}j} = \theta_{{\rm b}j} - \left(\theta_{{\rm v}j}+\frac{\pi}{2}\right) \,,
\end{eqnarray}
which implies that $\tan\theta_{{\rm b}j} \tan(\theta_{{\rm v}j}+\theta_{{\rm bv}j}) = -1$, and using the fact that
\begin{eqnarray}
\tan(\theta_1+\theta_2) = \frac{\tan\theta_1 + \tan\theta_2}{1-\tan\theta_1 \tan\theta_2} \,,
\end{eqnarray}
it is not difficult to show that 
\begin{eqnarray}
\label{eq:theta_bv}
\tan\theta_{{\rm bv}j} = -\mu = -\left(\frac{\nu-\eta}{2}\right)k^2 \,.
\end{eqnarray}
This means that $\mu\ne 0$ provides a measure of how non-orthogonal
velocity and magnetic field perturbations are.

It is evident that when the magnetic Prandtl number approaches unity
viscous, resistive, MRI-driven magnetic and velocity perturbations tend
to be orthogonal, i.e., $\tan \theta_{{\rm v}j}\tan \theta_{{\rm
    b}j}=-1$, and therefore
\begin{eqnarray}
\label{eq:theta_diff_Pmeq1}
  \theta_{\rm diff} \equiv \theta_{{\rm b}j} - \theta_{{\rm v}j} = \frac{\pi}{2} \quad
  \textrm{for} \quad {\rm Pm}=1 \,,
\end{eqnarray}
for every wavenumber $k$. This is illustrated in
Figure~\ref{fig:geom_def} which shows the evolution of the angles
$\theta_{{\rm b}}$ and $\theta_{{\rm v}}$ corresponding to the most
unstable MRI mode as a function of the Reynolds/magnetic Reynolds
number when the magnetic Prandtl number is equal to unity.

\section{Physical Structure of MRI Modes}
\label{sec:mri_modes}

The evolution of the physical structure of a single growing MRI mode
with wavenumber $0<k<k_{\rm c}$ is characterized by its growth rate
$\gamma_+$, the relative magnitude between the amplitudes of magnetic
and velocity field perturbations, $b_0/v_0$, and the two angles
defining the planes containing them, $\theta_{\rm b}$ and $\theta_{\rm
  v}$. For any reasonable spectrum of initial perturbations the mode
that exhibits the fastest exponential growth, $\gamma_{\rm max}$,
which we refer to as $k_{\rm max}$, will dominate the dynamics of the
late time evolution of the viscous, resistive MRI. It is therefore of
particular interest to characterize the physical properties of this
fastest growing mode in different dissipative regimes.

\subsection{ Marginal and Fastest Growing MRI-modes}
\label{sub:modes_marginal_fastest}

Because the eigenvalue associated with the unstable growing mode,
$\gamma_+$, is always real for any combination of the Reynolds and
magnetic Reynolds numbers, it is possible to find the marginally
stable mode $k_{\rm c}$ such that $\gamma_+(k_{\rm c}) \equiv 0$.
Setting $\sigma=0$ in equation~(\ref{eq:dispersion_relation_nu_eta}),
we obtain a polynomial in $k_{\rm c}$
\begin{eqnarray}
\label{eq:k_c}
k_{\rm c}^2(1 + \nu \eta k_{\rm c}^2)^2 + \kappa^2(1 + \eta^2 k_{\rm c}^2) - 4  = 0 \,,
\end{eqnarray}
valid for any value of the viscosity and resistivity.  Note that
$k_{\rm c}$ sets the minimum domain height for numerical simulations
of viscous, resistive MRI-driven turbulence.  Figure~\ref{fig:k_c}
shows the solutions of equation (\ref{eq:k_c}) in various dissipative
regimes for Keplerian rotation.  The analytic solutions of equation
(\ref{eq:k_c}) are algebraically complicated but their asymptotic
limits are rather simple. We find expressions for this critical
wavenumber in several regimes of interest below.

In the ideal MHD limit, it is straightforward to find simple analytical
expressions for the most unstable wavenumber, $k_{\rm max}$, and its
associated growth rate, $\gamma_{\rm max}$.  However, the analytical
expressions that we derived for the eigenfrequencies in the non-ideal
case, equation (\ref{eq:eigenvalues_all}), are not amenable to the
usual extremization procedure. More precisely, it is very challenging
to find the values of $k_{\rm max}$ and $\gamma_{\rm max}$ that
satisfy
\begin{equation}
  \left.\frac{d\gamma_+}{dk}\right|_{k_{\rm max}}=0 \,.
\end{equation}
Figures~\ref{fig:k_max}~and~\ref{fig:gamma_max} show the solutions of
this equation in various dissipative regimes.

Another possible path to find the values of the wavenumber $k_{\rm
  max}$, and the associated growth rate, is to use the fact that
$\gamma_{\rm max}$ satisfies simultaneously the dispersion relation
(\ref{eq:dispersion_relation_nu_eta}) and its derivative to eliminate
$k_{\rm max}$ between these two and obtain a polynomial in
$\gamma_{\rm max}$. The largest of the roots of this polynomial is the
desired maximum growth rate.  It is possible to find $k_{\rm max}$
following a similar methodology, but eliminating between the two
polynomials $\gamma_{\rm max}$ instead.  However, for arbitrary values
of the viscosity and resistivity, both procedures lead to a seventh
degree polynomial whose roots \emph{must} be found numerically,
defeating altogether the attempt to find analytical expressions for
$k_{\rm max}$ and $\gamma_{\rm max}$.

Using as a guide the results shown in
Figures~\ref{fig:k_max}~and~\ref{fig:gamma_max}, we follow an
alternative procedure. The goal is to find simple analytical
expressions to describe the asymptotic behavior of the most unstable
mode, $k_{\rm max}$, and the maximum growth rate, $\gamma_{\rm max}$,
in different dissipative regimes.  It is evident from
Figure~\ref{fig:growths} that $k_{\rm max}<1$ and $\gamma_{\rm max}<1$
for all the non-ideal MRI modes.  This information can be used to
simplify the dispersion relation and its derivative so as to decrease
their order without loosing vital information. This makes it possible
to obtain manageable, but accurate, expressions for $k_{\rm max}$ and
$\gamma_{\rm max}$ in different limiting regimes.

\begin{figure*}[t]
  \includegraphics[width=0.675\columnwidth,trim=0 0 0 0]{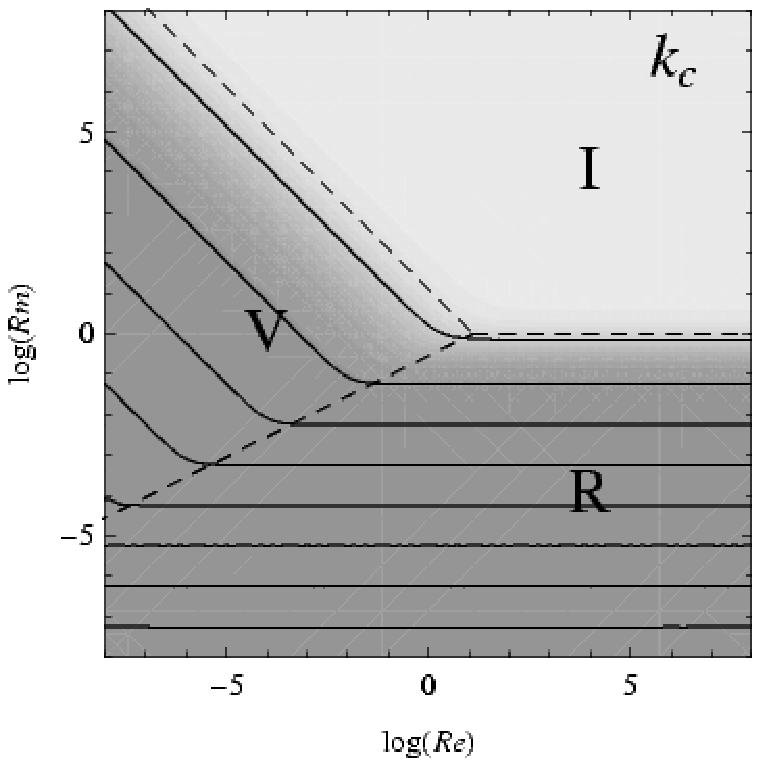}
  \includegraphics[width=0.675\columnwidth,trim=0 0 0 0]{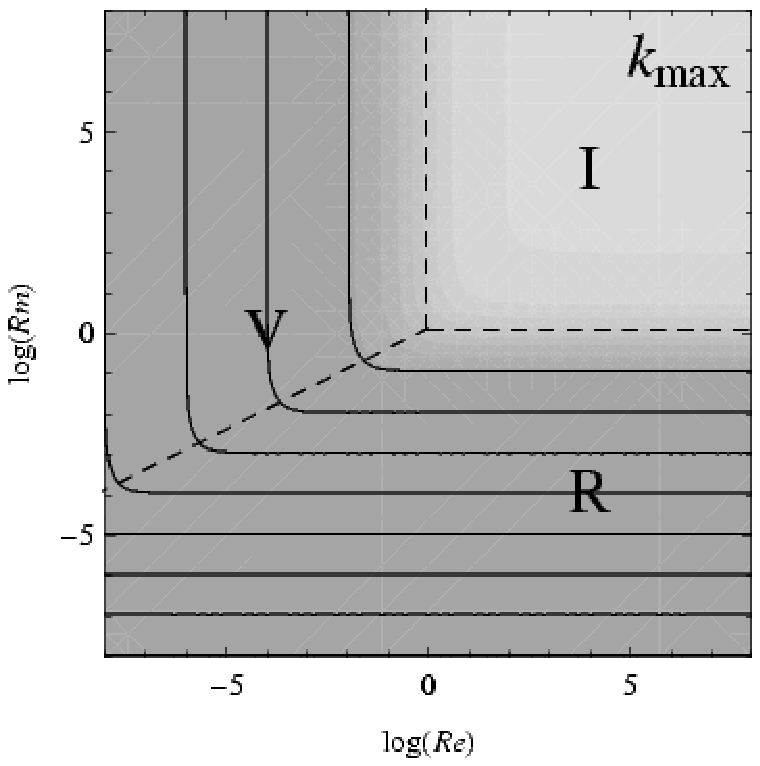}
  \includegraphics[width=0.675\columnwidth,trim=0 0 0 0]{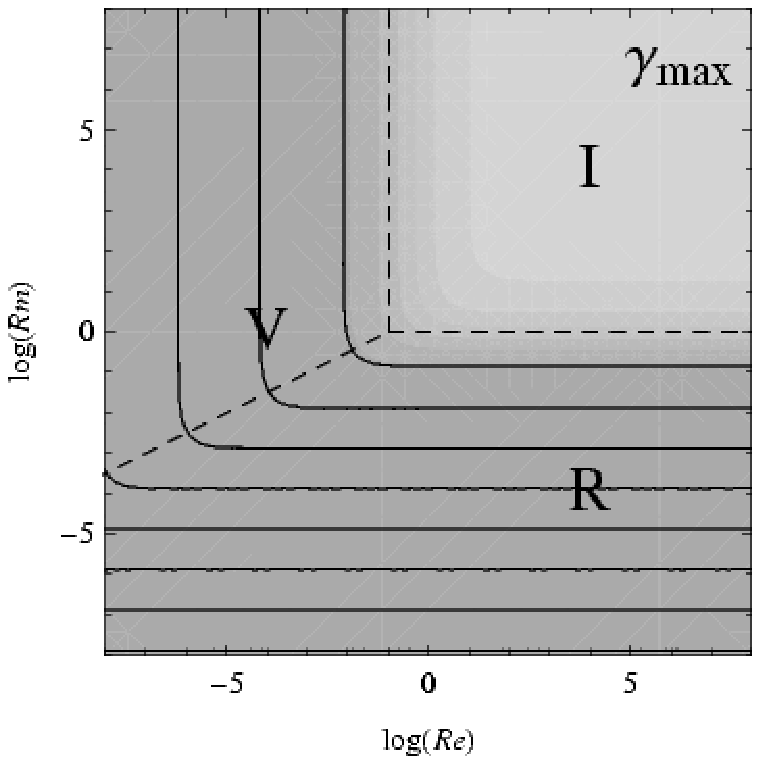}
  \caption{Contour plots for the critical wavenumber, $k_{\rm c}$, the
    most unstable wavenumber, $k_{\rm max}$, and the maximum growth
    rate, $\gamma_{\rm max}$, for Keplerian rotation. In all three
    panels, lighter gray areas correspond to larger values of $k_{\rm
      c}$, $k_{\rm max}$, and $\gamma_{\rm max}$, respectively.  The
    solid lines highlight the contours for $k_{\rm c} = 1, \ldots,
    10^{-7}$ and $k_{\rm max} = 10^{-1}, \dots, 10^{-8}$.  The labels I
    (ideal), R (resistive), and V (viscous), denote the three regions
    of the $({\rm Re}, {\rm Rm})$ plane where equations
    (\ref{eq:k_c_ideal}), (\ref{eq:k_max_ideal}), and
    (\ref{eq:gamma_max_ideal}); (\ref{eq:k_c_lim_reggrm}),
    (\ref{eq:k_max_lim_reggrm}) and (\ref{eq:gamma_max_lim_reggrm});
    and (\ref{eq:k_c_lim_rellrm}), (\ref{eq:k_max_lim_rellrm}), and
    (\ref{eq:gamma_max_lim_rellrm}) are valid, respectively.  The
    dashed lines dividing the three regions are obtained by equating
    neighboring approximations for $k_{\rm c}$, $k_{\rm max}$, and
    $\gamma_{\rm max}$.}
  \label{fig:contours}
\end{figure*}

Figure~\ref{fig:contours} shows contour plots for the critical
wavenumber, $k_{\rm c}$, the most unstable wavenumber, $k_{\rm max}$,
and the maximum growth rate, $\gamma_{\rm max}$, as a function of the
Reynolds and magnetic Reynolds numbers for Keplerian rotation.  In all
three panels, lighter gray areas correspond to larger values of
$k_{\rm c}$, $k_{\rm max}$, and $\gamma_{\rm max}$, respectively.
Note that in all the cases, the functional form of the contours
naturally divides the plane $({\rm Re}, {\rm Rm})$ in three
distinctive regions that we denote according to ${\rm I}$ (ideal),
${\rm R}$ (resistive), and ${\rm V}$ (viscous).  Note that when the
most unstable wavenumber, $k_{\rm max}$, and the maximum growth rate,
$\gamma_{\rm max}$, are considered, these regions can be associated
with the regions where ${\rm Re, \,Rm} \gg 1$, ${\rm Re} \gg {\rm
  Rm}$, and ${\rm Re} \ll {\rm Rm}$, respectively.  The overlap
between these regions is not as clear when the critical wavenumber
$k_{\rm c}$ is considered and some care is needed when deriving
approximated expressions for it.

\subsection{Ideal MRI Modes}
\label{sub:modes_ideal}

Let us first demonstrate briefly how the formalism presented in
\S~\ref{sec:solution} reduces to previously known results in the ideal
MHD limit. In the absence of dissipation, the eigenvalues
$\{\sigma_{0,j}\}$, with $j=1,2,3,4$, are the roots of the dispersion
relation associated with the ideal MRI \citep{BH91,BH98},
\begin{equation}
\label{eq:dispersion_relation_ideal}
(k^2+\sigma_{0,j}^2)^2 + 
\kappa^2( k^2  + \sigma_{0,j}^2) - 4 k^2   = 0 \,,
\end{equation}
and are given by \citep{PCP06}
\begin{equation}
\sigma_{0,j} = \pm \left(- \Lambda_0 \pm \sqrt{\Delta_0}\,\right)^{1/2} \,,
\end{equation}
where we have defined the quantities $\Lambda_0$ and $\Delta_0$ such that
\begin{eqnarray}
\label{eq:Lambda0}
\Lambda_0 &\equiv& \frac{\kappa^2}{2} + k^2  \,, \\ 
\label{eq:Delta0}
\Delta_0 &\equiv&  \frac{\kappa^4}{4} + 4 k^2   \,.
\end{eqnarray}

\begin{figure*}[t]
  \includegraphics[width=0.675\columnwidth,trim=0 0 0 0]{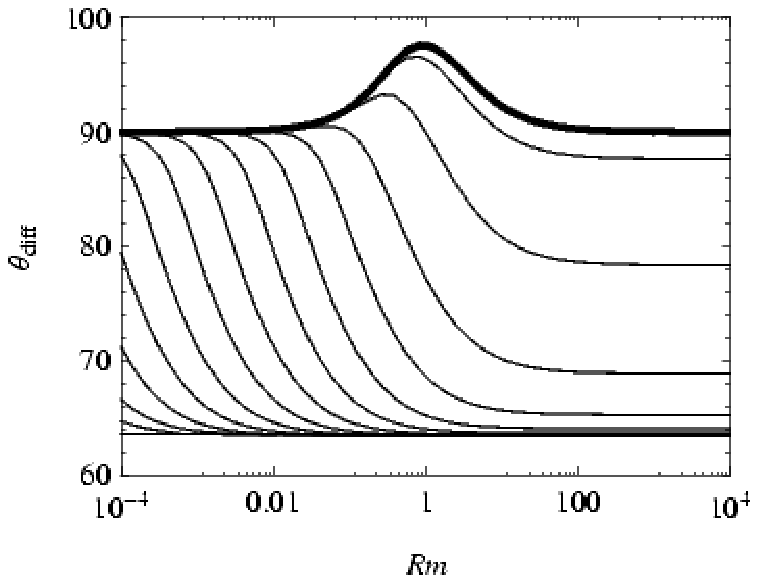}
  \includegraphics[width=0.675\columnwidth,trim=0 0 0 0]{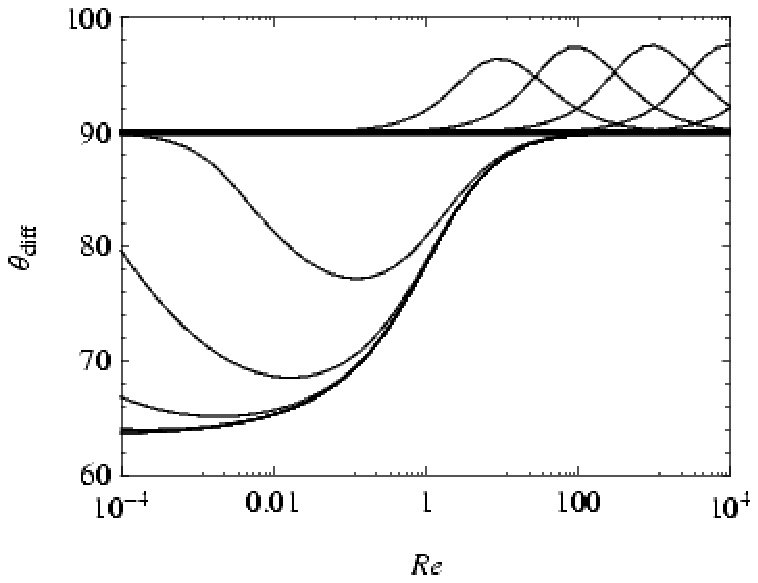}
  \includegraphics[width=0.675\columnwidth,trim=0 0 0 0]{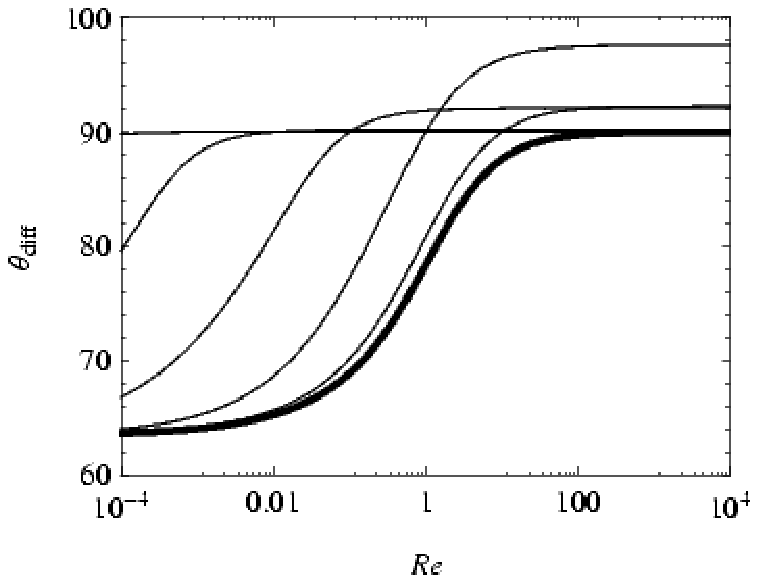}
  \caption{Opening angle, $\theta_{\rm diff}=\theta_{\rm
      b}-\theta_{\rm v}$, between the planes containing the fastest
    exponentially growing magnetic and velocity perturbations for
    Keplerian rotation in various dissipative regimes.  In the ideal
    MHD limit the opening angle is $\theta_{\rm diff}=\pi/2$, see
    eqs.~(\ref{eq:theta_v_ideal})~and~(\ref{eq:theta_b_ideal}).
    \emph{Left}: opening angle $\theta_{\rm diff}$ as a function of
    the magnetic Reynolds number for different values of the Reynolds
    number.  The thick solid line denotes the inviscid limit, i.e.,
    ${\rm Re}\rightarrow \infty$.  The thin solid lines, in decreasing
    order according to $\theta_{\rm diff}$ for fixed ${\rm Rm}$,
    correspond to ${\rm Re} =10, 1, \ldots, 10^{-12}$.  For magnetic
    Reynolds numbers much larger than unity, the opening angle is
    independent of ${\rm Rm}$ regardless of the value of ${\rm Re}$.
    For sufficiently small/large ${\rm Rm}$, $\theta_{\rm
      diff}\rightarrow \pi/2$ provided that ${\rm Re}\gg {\rm Rm}$.
    For sufficiently small/large ${\rm Rm}$, $\theta_{\rm
      diff}\rightarrow \pi/2-\arctan(\kappa/2)$ provided that ${\rm
      Rm}\ll {\rm Re}$. Note that this corresponds to $\theta_{\rm
      diff}=63\degr 26'$ for a Keplerian disk. The only conditions
    under which $\theta_{\rm diff}$ exceeds $\pi/2$ are such that
    ${\rm Re} \ge {\rm Rm} \simeq 1$.  \emph{Middle}: opening angle
    $\theta_{\rm diff}$ as a function of the Reynolds number for
    different values of the magnetic Prandtl number. The thick solid
    line at $\theta_{\rm diff}=\pi/2$ corresponds to ${\rm Pm}=1$, see
    eq.~ (\ref{eq:theta_diff_Pmeq1}) and Fig.~\ref{fig:geom_def}.  The
    thin solid lines with peaks at $\theta_{\rm diff}>\pi/2$
    correspond, from left to right, to ${\rm Pm}=10^{-1}, 10^{-2},
    \ldots$.  The thin solid lines, with $\theta_{\rm diff}<\pi/2$, in
    decreasing order according to $\theta_{\rm diff}$ for fixed ${\rm
      Re}$, correspond to ${\rm Pm} =10, 10^2, \ldots$.  Note that for
    small ${\rm Re}$, $\theta_{\rm diff}\rightarrow \pi/2$ for ${\rm
      Pm}\simeq 1$, while $\theta_{\rm diff}\rightarrow
    \pi/2-\arctan(\kappa/2)$ for ${\rm Pm}\gg 1$. \emph{Right}:
    opening angle $\theta_{\rm diff}$ as a function of the Reynolds
    number for different values of the magnetic Reynolds number.  The
    thick solid line corresponds to the ideal conductor limit, i.e.,
    ${\rm Rm}\rightarrow \infty$.  The thin solid lines, from right to
    left, correspond to ${\rm Rm} =10, \ldots, 10^{-3}$.  For Reynolds
    numbers larger than unity, the opening angle is independent of
    ${\rm Re}$ regardless of the value of ${\rm Rm}$.}
  \label{fig:theta_diff}
\end{figure*}

The critical wavenumber for the onset of the ideal MRI is obtained by
setting $\sigma_0=0$ in the dispersion relation
(\ref{eq:dispersion_relation_ideal}), this leads to
\begin{equation}
\label{eq:k_c_ideal}
k_{\rm c}  = \sqrt{2q} = \sqrt{4-\kappa^2} \,.
\end{equation}
For all the modes with wavenumbers $k<k_{\rm c}$ the difference
$\sqrt{\Delta_0}-\Lambda_0$ is positive and we can define the ``growth
rate'' $\gamma_0$ and the ``oscillation frequency'' $\omega_0$ by
\begin{eqnarray}
\label{eq:gamma}
\gamma_0   &\equiv& \left(\sqrt{\Delta_0} -\Lambda_0\right)^{1/2} \,, \\
\label{eq:omega}
\omega_0 &\equiv&  \left(\sqrt{\Delta_0} + \Lambda_0\right)^{1/2} \,,
\end{eqnarray}
both of which are real and positive (for all positive values of the
parameter $q$). This shows that two of the solutions of equation
(\ref{eq:dispersion_relation_ideal}) are real and the other two are
imaginary.  We can thus write the four eigenvalues in compact notation
as
\begin{eqnarray}
\label{eq:eigenvalues}
\sigma_{0,1} = \gamma_0 \,, \quad
\sigma_{0,2} = -\gamma_0 \,,\quad
\sigma_{0,3} = i \omega_0 \,, \quad
\sigma_{0,4} = -i \omega_0 \,. \nonumber \\
\end{eqnarray}

In the ideal MHD limit, it is evident that the velocity and magnetic
field perturbations are orthogonal for any mode, i.e., $\tan
\theta_{{\rm v}} \tan \theta_{{\rm b}}=-1$, see
equation~(\ref{eq:theta_bv}), and therefore
\begin{eqnarray}
  \theta_{{\rm b}} = \theta_{{\rm v}} + \frac{\pi}{2} \,.
\end{eqnarray}
The temporal evolution of a single MRI-unstable mode in physical space
reduces to
\begin{equation}
\bb{\delta}(z, t) = 
\frac{\sqrt{2} \, e^{\gamma_0 t}}{\sqrt{v_0^2+b_0^2}} \,
  \left[\begin{array}{r}
     v_0 \cos\theta_{{\rm v}} \sin(kz) \\ 
     v_0 \sin\theta_{{\rm v}} \sin(kz) \\ 
     b_0 \sin\theta_{{\rm v}} \cos(kz) \\ 
   - b_0 \cos\theta_{{\rm v}} \cos(kz)
  \end{array}\right] \,.
\end{equation}
These are essentially the (normalized) perturbations found in equation
(4) in \citet{GX94} \footnote{Note that the angle $\gamma$ in
  \citet{GX94}, in our notation defined by $\tan \gamma = -\delta
  B_r/\delta B_\phi$, is such that $\gamma=0$ in the positive
  azimuthal axis and it takes increasingly positive values in the
  counter-clockwise direction.}.

From the definition of the angle $\theta_{{\rm v}}$, see equation
(\ref{eq:theta_v}), it can be seen that 

The maximum growth rate can be obtained by noting that $\gamma_{0} = q
\sin\theta_{{\rm v}} \cos\theta_{{\rm v}}$ and therefore the maximum
growth corresponds to
\begin{equation}
\label{eq:gamma_max_ideal}
\gamma_{\rm max} = \frac{q}{2} = 1 - \frac{\kappa^2}{4}
\,.
\end{equation}
It then follows that, in the absence of dissipation, the planes
containing the exponentially growing velocity and magnetic field
perturbations are characterized by the angles
\begin{eqnarray}
\label{eq:theta_v_ideal}
\theta_{{\rm v}} &=& \frac{\pi}{4} \,, \\
\label{eq:theta_b_ideal}
\theta_{{\rm b}} &=& \frac{3\pi}{4} \,,
\end{eqnarray}
regardless of the value of the shearing parameter/epicyclic frequency.

Finally, noting that the wavenumber for which the maximum growth rate
is realized is
\begin{equation}
\label{eq:k_max_ideal}
k_{\rm max} = \sqrt{1-\frac{\kappa^4}{16}} \,,
\end{equation}
and using equation~(\ref{eq:b0_v0}) for the ratio between the
amplitudes of the magnetic and velocity fields we obtain
\begin{equation}
\label{eq:b0_v0_ideal}
\frac{b_0}{v_0} =
\sqrt{\frac{4+\kappa^2}{4-\kappa^2}}\,.
\end{equation}

In \S~\ref{sub:MRI_energetics} we derive equations for the MRI-driven
Reynolds and Maxwell stresses, as well as the kinetic and magnetic
energy densities associated with the perturbations.  Equations
(\ref{eq:mean_Rrphi}), (\ref{eq:mean_Mrphi}), (\ref{eq:mean_EK}), and
(\ref{eq:mean_EM}), show why, in the ideal MHD limit, equations
(\ref{eq:theta_v_ideal}), (\ref{eq:theta_b_ideal}), and
(\ref{eq:b0_v0_ideal}) are the reason for which the ratio between the
Maxwell to the Reynolds stresses is identical to the ratio between
magnetic and kinetic energy densities for any shear parameter and
equal to $5/3$ in the Keplerian case \citep{PCP06}.

\begin{figure*}[t]
  \includegraphics[width=0.675\columnwidth,trim=0 0 0 0]{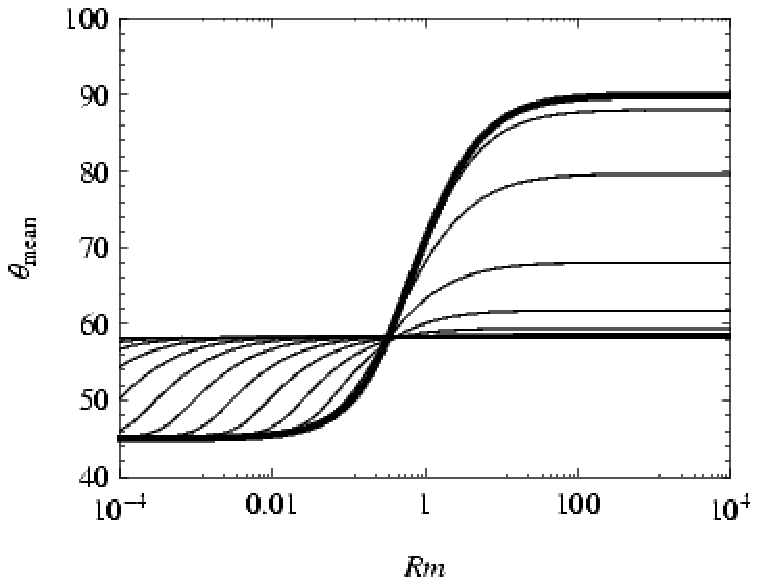}
  \includegraphics[width=0.675\columnwidth,trim=0 0 0 0]{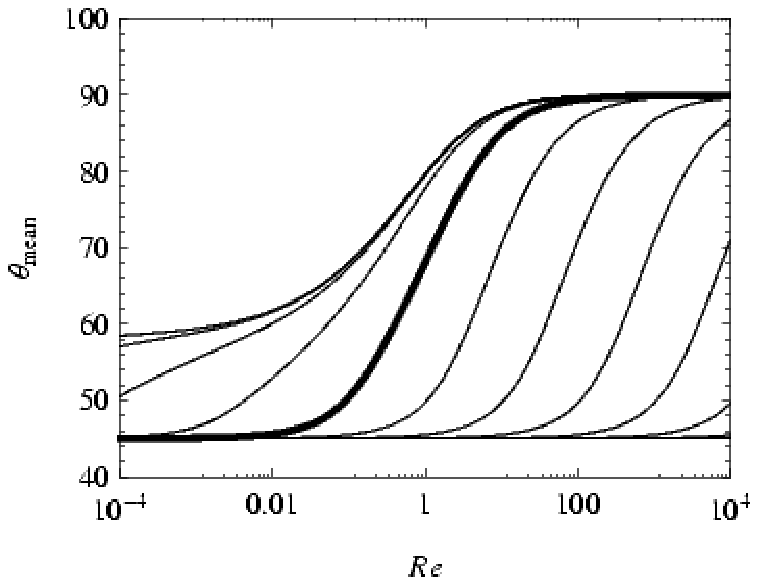}
  \includegraphics[width=0.675\columnwidth,trim=0 0 0 0]{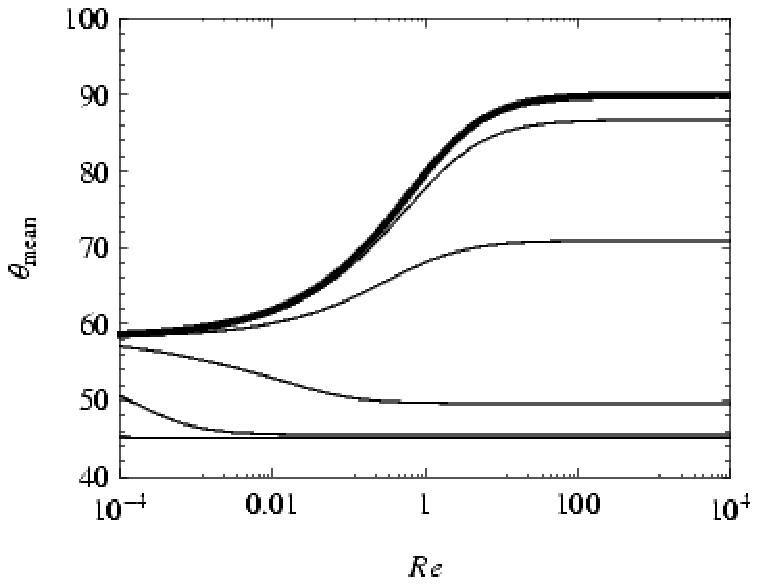}
  \caption{Mean angle, $\theta_{\rm mean}=(\theta_{\rm v}+\theta_{\rm
      b})/2$, defined by the fastest exponentially growing velocity
    and magnetic perturbations for Keplerian rotation in various
    dissipative regimes. Note that $\theta_{\rm mean}=\pi/2$ in the
    ideal MHD limit, see
    eqs.~(\ref{eq:theta_v_ideal})~and~(\ref{eq:theta_b_ideal}).
    \emph{Left}: mean angle $\theta_{\rm mean}$ as a function of the
    magnetic Reynolds number for different values of the Reynolds
    number.  The thick solid line denotes the inviscid limit, i.e.,
    ${\rm Re}\rightarrow \infty$.  There is a critical magnetic
    Reynolds number ${\rm Rm}_{\rm c} \lesssim 1$ that differentiates
    the asymptotic limits of $\theta_{\rm mean}$ for large and small
    Reynolds numbers.  When the Reynolds number changes from ${\rm
      Re}\ll {\rm Rm}$ to ${\rm Re}\gg {\rm Rm}$, the mean angle
    evolves according to $\theta_{\rm
      mean}:[\pi/2+\arctan(\kappa/2)]/2 \rightarrow \pi/2$ for ${\rm
      Rm}>{\rm Rm}_{\rm c}$, while $\theta_{\rm
      mean}:[\pi/2+\arctan(\kappa/2)]/2 \rightarrow \pi/4$ for ${\rm
      Rm}<{\rm Rm}_{\rm c}$. Note that for Keplerian rotation
    $[\pi/2+\arctan(\kappa/2)]/2=58\degr 17'$. \emph{Middle}: opening
    angle $\theta_{\rm diff}$ as a function of the Reynolds number for
    different values of the magnetic Prandtl number.  The thick solid
    line corresponds to ${\rm Pm}=1$. The thin solid lines to the
    right correspond to smaller values ${\rm Pm}=10^{-1}, 10^{-2},
    \ldots$. The thin solid lines to the left correspond to ${\rm Pm}
    =10, 10^2, \ldots$.  Note that for small ${\rm Re}$, $\theta_{\rm
      mean}\rightarrow \pi/4$ for ${\rm Pm}\lesssim 1$, while
    $\theta_{\rm mean}\rightarrow [\pi/2+\arctan(\kappa/2)]/2$ for
    ${\rm Pm}\gg 1$.  \emph{Right}: mean angle $\theta_{\rm mean}$ as
    a function of the magnetic Reynolds number for different values of
    the Reynolds number.  The thick solid line corresponds to the
    ideal conductor limit, i.e., ${\rm Rm}\rightarrow \infty$.  The
    thin solid lines, in decreasing order, correspond to ${\rm Re}
    =10, 1, \ldots$.  For Reynolds numbers larger than unity, the mean
    angle is independent of ${\rm Re}$ regardless of the value of
    ${\rm Rm}$.}
  \label{fig:theta_mean}
\end{figure*}

\subsection{MRI Modes with ${\rm Re} \gg {\rm Rm}$}
\label{sub:modes_re_gg_rm}

Lets us consider the inviscid, poorly conducting limit described by
$\nu=0$ and $\eta\gg 1$.  In this case, the marginally stable mode
satisfying equation (\ref{eq:k_c}) is given by
\begin{eqnarray}
\label{eq:k_c_lim_reggrm}
k_{\rm c}  = \sqrt{\frac{4-\kappa^2}{1+\eta^2\kappa^2}} \,.
\end{eqnarray}
The dependence of this critical wavenumber on the magnetic Reynolds
number is shown on the left panel in Figure~\ref{fig:k_c}, which shows
that for small magnetic Reynolds numbers $k_{\rm c} \propto Rm$.

As discussed in \S~\ref{sub:modes_marginal_fastest}, finding an
analytic expression for the maximum growth rate and wavenumber
associated with it is not as straightforward. The left panel of
Figure~\ref{fig:gamma_max} suggests that in the limit ${\rm
  Re}\rightarrow \infty$ and ${\rm Rm}\ll 1$, the maximum growth rate
is linear in the magnetic Reynolds number, $\gamma_{\rm max} \propto
{\rm Rm} \propto \eta^{-1}$. This information can be used to derive
asymptotic expressions for the dispersion relation
(\ref{eq:dispersion_relation_nu_eta}) and its derivative.  The leading
order contributions are given by
\begin{eqnarray}
\label{eq:disp_lim_reggrm}
\kappa^2 \gamma_{\rm max}^2 + 2\kappa^2 \eta k_{\rm max}^2\gamma_{\rm max} +
\kappa^2\eta^2k_{\rm max}^4 +(\kappa^2-4)k_{\rm max}^2= 0 \,,
\nonumber \\
\end{eqnarray}
and
\begin{eqnarray}
\label{eq:disp_lim_reggrm_deriv}
2\kappa^2 \eta \gamma_{\rm max} +
2\kappa^2\eta^2k_{\rm max}^2 + \kappa^2-4= 0 \,,
\end{eqnarray}
respectively.

Eliminating either $\gamma_{\rm max}$ or $k_{\rm max}$ between
equations (\ref{eq:disp_lim_reggrm}) and
(\ref{eq:disp_lim_reggrm_deriv}) we obtain
\begin{eqnarray}
\label{eq:k_max_lim_reggrm}
k_{\rm max} = \frac{1}{\eta}\sqrt{\frac{4-\kappa^2}{4\kappa^2}} \,,
\end{eqnarray}
and
\begin{eqnarray}
\label{eq:gamma_max_lim_reggrm}
\gamma_{\rm max} = \frac{1}{\eta}\frac{4-\kappa^2}{4\kappa^2} \,.
\end{eqnarray}
In this case, $\gamma_{\rm max}= \eta k_{\rm max}^2$ for any value of
the epicyclic frequency $\kappa$.

The dependence of both $k_{\rm max}$ and $\gamma_{\rm max}$ in this
limiting case is shown with dashed lines in the left panels of
Figures~\ref{fig:k_max} and \ref{fig:gamma_max}, respectively. The
agreement between equations (\ref{eq:k_max_lim_reggrm}) and
(\ref{eq:gamma_max_lim_reggrm}) and the solutions to the full
dispersion relation (\ref{eq:dispersion_relation_nu_eta}) in the case
$\nu=0$ and $\eta\gg1$ is excellent, only breaking down close to
magnetic Reynolds numbers of order unity. Note that even though the
equations (\ref{eq:k_max_lim_reggrm}) and
(\ref{eq:gamma_max_lim_reggrm}) were derived under the assumption of
an inviscid fluid, i.e., $\nu=0$, these expressions can describe the
asymptotic behavior of both $k_{\rm max}$ and $\gamma_{\rm max}$ for
finite Reynolds numbers provided that the conditions ${\rm Re} \gg
{\rm Rm}$ and ${\rm Rm}\ll 1$ are satisfied.

Substituting the asymptotic expressions for $k_{\rm max}$ and
$\gamma_{\rm max}$ in equations (\ref{eq:k_max_lim_reggrm}) and
(\ref{eq:gamma_max_lim_reggrm}) into equation (\ref{eq:b0_v0}) we obtain
the ratio between the amplitudes of the magnetic and velocity field
perturbations
\begin{equation}
\label{eq:b0_v0_reggrm}
  \frac{b_0}{v_0} =
  \frac{\eta\kappa^3}{\sqrt{4-\kappa^2}}\,.
\end{equation}
Therefore, inviscid, resistive MRI-unstable modes are dominated by
magnetic field perturbations. Note that the ratio between amplitudes
increases linearly with resistivity.

The asymptotic behavior for the angles characterizing velocity and
magnetic field perturbations, equations (\ref{eq:theta_v}) and
(\ref{eq:theta_b}), are given by
\begin{eqnarray}
\tan \theta_{\rm v} = \frac{1}{2\kappa^2 \eta} \,,
\end{eqnarray}
and
\begin{eqnarray}
\tan \theta_{\rm b} = - \eta \kappa^2 \,.
\end{eqnarray}
In the limit ${\rm Re}\rightarrow\infty$ and ${\rm Rm}\rightarrow 0$, we obtain
\begin{eqnarray}
\lim_{\eta \rightarrow \infty} \theta_{\rm v} &=& 0 \,, \\
\lim_{\eta \rightarrow \infty} \theta_{\rm b} &=& \frac{\pi}{2} \,.
\end{eqnarray}

\begin{figure*}[t]
  \includegraphics[width=2\columnwidth,trim=0 0 0 0]{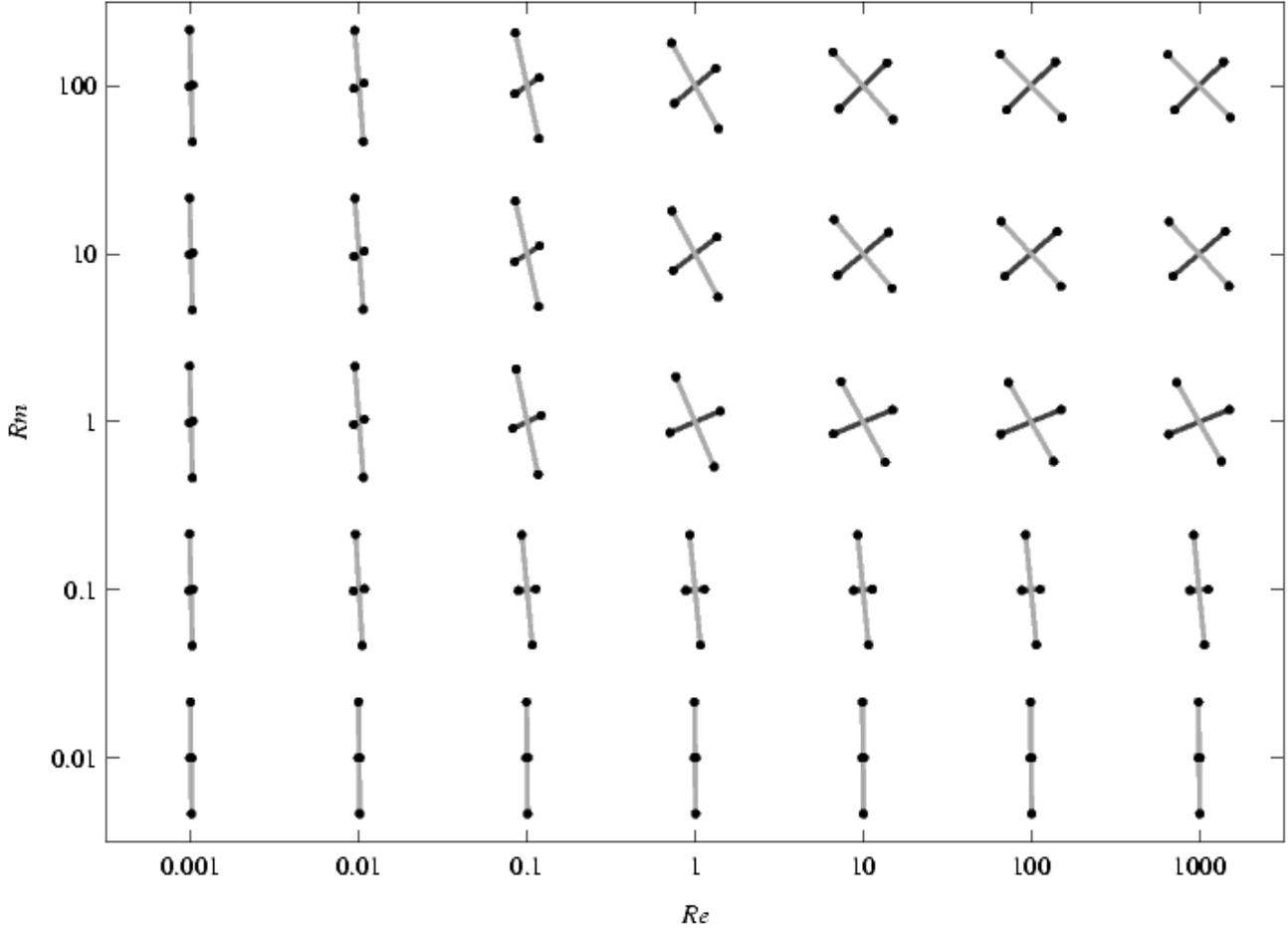}
  \caption{Geometrical representation of viscous, resistive MRI modes
    for varying Reynolds and magnetic Reynolds numbers. The black/gray
    lines denote velocity/magnetic components of the most unstable
    mode.}
  \label{fig:geom_sketch}
\end{figure*}

We therefore conclude that in the regime of large Reynolds numbers and
small magnetic Reynolds numbers, magnetic field perturbations are
larger than velocity field perturbations, both fields tend to be
orthogonal and aligned with the azimuthal and radial directions,
respectively, see
Figures~\ref{fig:theta_diff},~\ref{fig:theta_mean},~and~\ref{fig:geom_sketch}.

%\newpage

\subsection{MRI Modes with ${\rm Re}={\rm Rm}\ll 1$}
\label{sub:modes_re_eq_rm_ll_1}

When the magnetic Prandtl number is unity, and with $\eta=\nu\gg 1$,
the marginally stable mode satisfying equation (\ref{eq:k_c}) is given
by
\begin{eqnarray}
\label{eq:k_c_lim_reeqrm}
k_{\rm c}  = \frac{\sqrt{4-\kappa^2}}{\eta\kappa} \,.
\end{eqnarray}
The dependence of this critical wavenumber on the Reynolds number is
shown in the middle panel in Figure~\ref{fig:k_c}. Incidentally,
equation~(\ref{eq:k_c_lim_reeqrm}) corresponds to the asymptotic limit
$\eta\gg 1$ of equation~(\ref{eq:k_c_lim_reggrm}).

It is not hard to see that the leading order contributions to the
dispersion relation and its derivative in the limit ${\rm Re}={\rm
  Rm}\ll 1$ are identical to the ones obtained in the case ${\rm
  Re}\rightarrow \infty$ and ${\rm Rm}\ll 1$.  Therefore, all the
expressions for $k_{\rm max}$, $\gamma_{\rm max}$, $\theta_{\rm v}$,
and $\theta_{\rm b}$, derived in \S~\ref{sub:modes_re_gg_rm}, are also
valid in this case.  The dependence of both $k_{\rm max}$ and
$\gamma_{\rm max}$ in this limiting case is shown with dashed lines in
the middle panels of Figures~\ref{fig:k_max} and \ref{fig:gamma_max},
respectively. The agreement between equations
(\ref{eq:k_max_lim_rellrm}) and (\ref{eq:gamma_max_lim_rellrm}) and
the solutions to the full dispersion relation
(\ref{eq:dispersion_relation_nu_eta}) in the case $\nu=\eta$, i.e.,
${\rm Pm}=1$, is also excellent in this case.

\subsection{MRI Modes with ${\rm Re} \ll {\rm Rm}$}
\label{sub:modes_re_ll_rm}

Lets us consider next the highly viscous, ideal conductor limit
described by $\nu\gg1$ and $\eta=0$. In this case, the marginally
stable mode satisfying equation (\ref{eq:k_c}) is given by
\begin{eqnarray}
  \label{eq:k_c_lim_rellrm}
k_{\rm c}  = \left(\frac{\sqrt{4-\kappa^2}}{\nu\eta}\right)^{1/3} \,.
\end{eqnarray}

The right panel of Figure~\ref{fig:gamma_max} suggests that in the
limit ${\rm Re}\ll 1$ and ${\rm Rm}\rightarrow \infty$, the dependence
of the maximum growth on the Reynolds number is $\gamma_{\rm max}
\propto {\rm Re}^{1/2} \propto \nu^{-1/2}$. This information can be
used to derive asymptotic expressions for the dispersion relation
(\ref{eq:dispersion_relation_nu_eta}) and its derivative.  The leading
order contributions are given by
\begin{eqnarray}
\label{eq:disp_lim_rellrm}
(\kappa^2 +\nu^2k_{\rm max}^4)\gamma_{\rm max}^2 + (\kappa^2 -4)k_{\rm
  max}^2 =0 \,,
\end{eqnarray}
and
\begin{eqnarray}
\label{eq:disp_lim_rellrm_deriv}
2\nu^2k_{\rm max}^2 \gamma_{\rm max}^2 +\kappa^2-4= 0 \,,
\end{eqnarray}
respectively.

Eliminating either $\gamma_{\rm max}$ or  $k_{\rm max}$ between these
equations we obtain
\begin{eqnarray}
\label{eq:k_max_lim_rellrm}
k_{\rm max} = \sqrt{\frac{\kappa}{\nu}} \,,
\end{eqnarray}
and
\begin{eqnarray}
\label{eq:gamma_max_lim_rellrm}
\gamma_{\rm max} = \sqrt{\frac{4-\kappa^2}{2\nu\kappa}} \,.
\end{eqnarray}
Eliminating the epicyclic frequency between equations
(\ref{eq:gamma_max_lim_rellrm}) and (\ref{eq:k_max_lim_rellrm}) we
obtain
\begin{eqnarray}
\gamma_{\rm max}^2 = \frac{4-\nu^2k_{\rm max}^4}{2\nu^2k_{\rm max}^2} \,.
\end{eqnarray}

The dependence of both $k_{\rm max}$ and $\gamma_{\rm max}$ in this
limiting case is shown with dashed lines in the right panels of
Figures~\ref{fig:k_max} and \ref{fig:gamma_max}, respectively. The
agreement between equations (\ref{eq:k_max_lim_rellrm}) and
(\ref{eq:gamma_max_lim_rellrm}) and the solutions to the full dispersion
relation (\ref{eq:dispersion_relation_nu_eta}) in the case $\nu\gg 1$
and $\eta = 0$ is excellent, only breaking down close to Reynolds
numbers of order unity. Note that even though the equations
(\ref{eq:k_max_lim_rellrm}) and (\ref{eq:gamma_max_lim_rellrm}) were
derived under the assumption of a perfectly conducting fluid, i.e.,
$\eta=0$, these expressions can describe the asymptotic behavior of
both $k_{\rm max}$ and $\gamma_{\rm max}$ for finite Reynolds numbers
provided that the conditions ${\rm Re} \ll {\rm Rm}$ and ${\rm Re}\ll
1$ are satisfied.

Substituting the asymptotic expressions for $k_{\rm max}$ and
$\gamma_{\rm max}$ in equations (\ref{eq:k_max_lim_rellrm}) and
(\ref{eq:gamma_max_lim_rellrm}) into equation (\ref{eq:b0_v0}) we
obtain the relative amplitude of the magnetic and velocity field
perturbations
\begin{equation}
\label{eq:b0_v0_rellrm}
  \frac{b_0}{v_0} =
  2\sqrt{\frac{\nu\kappa^{3}}{4+\kappa^2}}\,.
\end{equation}
Therefore, viscous, conducting MRI-unstable modes are also dominated
by magnetic field perturbations. In this case, the ratio between
amplitudes increases only with the square root of the viscosity.

The asymptotic behavior for the angles characterizing velocity and
magnetic field perturbations, equations (\ref{eq:theta_v}) and
(\ref{eq:theta_b}), are given by
\begin{eqnarray}
\tan \theta_{\rm v} = \frac{\kappa}{2} +
\frac{4+\kappa^2}{\sqrt{2\nu\kappa(4-\kappa^2)}} \,,
\end{eqnarray}
and
\begin{eqnarray}
\tan \theta_{\rm b} = \sqrt{\frac{\nu(4-\kappa^2)}{2\kappa}} \,.
\end{eqnarray}
In the limit ${\rm Re}\rightarrow0$ and ${\rm Rm}\rightarrow \infty$, we obtain
\begin{eqnarray}
\lim_{\nu \rightarrow \infty} \theta_{\rm v} &=& \arctan \left(\frac{\kappa}{2}\right) \,,\\
\lim_{\nu \rightarrow \infty} \theta_{\rm b} &=& \frac{\pi}{2} \,.
\end{eqnarray}

For Keplerian rotation, the angle between the fastest growing velocity
field perturbation and the radial direction is given by $\theta_{\rm
  v} = \arctan(1/2)=26\degr 34'$. Therefore, the opening angle between
the planes containing velocity and magnetic field perturbations is
$\theta_{\rm diff} = \theta_{\rm b} - \theta_{\rm v}= 63\degr 26'$ and
their mean value is $\theta_{\rm mean} = (\theta_{\rm b} + \theta_{\rm
  v})/2 = 58\degr 17'$. The right panels of
Figures~\ref{fig:theta_diff} and \ref{fig:theta_mean} show that both
of these results are in perfect agreement with the asymptotic behavior
of the full solutions derived directly from the original dispersion
relation (\ref{eq:dispersion_relation_nu_eta}).

We therefore conclude that in the regime of small Reynolds numbers and
large magnetic Reynolds numbers, magnetic perturbations are larger
than velocity perturbations. In this case, however, the perturbed
magnetic and velocity fields are not orthogonal. The perturbed
magnetic field tends to be aligned with the azimuthal direction but
the velocity field perturbations do not tend to be aligned with the
radial direction. The angle between both fields is determined entirely
by the epicyclic frequency $\kappa$, see
Figures~\ref{fig:theta_diff},~\ref{fig:theta_mean},~and~\ref{fig:geom_sketch}.

\subsection{MRI Modes with ${\rm Re}={\rm Rm}\gg 1$}
\label{sub:modes_re_eq_rm_gg_1}

As we show in Appendix \ref{sec:appendix}, when the Reynolds and
magnetic Reynolds numbers are large enough the solutions to the
dispersion relation (\ref{eq:dispersion_relation_nu_eta}) tend
smoothly to the solutions of the dispersion relation
(\ref{eq:dispersion_relation_ideal}) for the idealized case.
Therefore, in this limit we recover all the expressions derived in
\S~\ref{sub:modes_ideal}.

\section{Physics of Maximally Unstable MRI Modes}
\label{sec:mri_physics}

We have shown that the expressions derived for the most unstable
wavenumber, $k_{\rm max}$, and its associated growth rate,
$\gamma_{\rm max}$, obtained from the simplified equations
(\ref{eq:disp_lim_reggrm}) and (\ref{eq:disp_lim_rellrm}) are good
approximations to the solutions obtained directly from the dispersion
relation~(\ref{eq:dispersion_relation_nu_eta}), in the limits ${\rm
  Re} \gg {\rm Rm}$ and ${\rm Re} \ll {\rm Rm}$, respectively.  We can
now identify the various terms in the original set of equations of
motion~(\ref{eq:vx})--(\ref{eq:by}) that lead to
equations~(\ref{eq:disp_lim_reggrm})~and~(\ref{eq:disp_lim_rellrm}).
This allows us to better understand the forces that act to destabilize
magnetized fluid elements.

For the sake of clarity we write the equations in this section with
physical dimensions. We represent the temporal derivatives with a dot
and the derivatives with respect to the vertical coordinate $z$ with a
prime.

\subsection{MRI Modes with ${\rm Re} \gg {\rm Rm}$}
\label{sub:mri_physics_re_gg_rm}

The equations of motion that lead to the dispersion relation
(\ref{eq:disp_lim_reggrm}) are given by
\begin{eqnarray}
\label{eq:vx_reggrm}
\dot{\delta v_r} &=& 2 \Omega_0 \delta v_\phi + \frac{\bar
  B_z}{4\pi\rho} \,\delta B_r' \,,  \\
\label{eq:vy_reggrm}
0 &=& - (2-q)\Omega_0 \delta v_r + \frac{\bar B_z}{4\pi\rho} \,\delta B_\phi' \,, \\
\label{eq:bx_reggrm}
\dot{\delta B_r} &=&  \bar B_z \delta v_r' + \eta \, \delta B_r'' \,,  \\
\label{eq:by_reggrm}
\dot{\delta B_\phi} &=& - q \Omega_0 \delta B_r + \eta \,  \delta B_\phi''\,.
\end{eqnarray}
Therefore, maximally unstable modes with ${\rm Re} \gg {\rm Rm}$ are
characterized by motions with radial acceleration due to the Coriolis
force acting on azimuthally displaced fluid elements and magnetic
tension.  Azimuthal force balance is attained via the joint action of
the Coriolis force acting on radially displaced fluid elements, radial
advection of background flow, and magnetic tension in the azimuthal
direction. The rate of change of the radial magnetic field
perturbations is due to the creation of radial field out of the
vertical background frozen into the radial velocity field with a
vertical gradient and field diffusion.  The rate of change of the
azimuthal magnetic field perturbations is due to the shearing of
radial magnetic field perturbations and field diffusion.

\subsection{MRI Modes with ${\rm Re} \ll {\rm Rm}$}
\label{sub:mri_physics_re_ll_rm}

The set of equations that lead to the dispersion relation
(\ref{eq:disp_lim_rellrm}) are given by
\begin{eqnarray}
\label{eq:vx_rellrm}
0 &=& 2 \Omega_0 \delta v_\phi +\nu  \delta v_r'' \,,  \\
\label{eq:vy_rellrm}
0 &=& - (2-q)\Omega_0 \delta v_r + \frac{\bar B_z}{4\pi\rho} \,\delta B_\phi' + \nu \delta v_\phi'' \,, \\
\label{eq:bx_rellrm}
\dot{\delta B_r} &=&  \bar B_z \delta v_r' \,,  \\
\label{eq:by_rellrm}
\dot{\delta B_\phi} &=& - q \Omega_0 \delta B_r \,.
\end{eqnarray}
In this case, maximally unstable modes with ${\rm Re} \ll {\rm Rm}$
are characterized by fluid displacements that take place under force
balance in both the radial and azimuthal directions.  The Coriolis
force acting on azimuthally displaced fluid elements is balanced by
the viscous force in the radial direction.  Azimuthal force balance is
attained via the joint action of the Coriolis force acting on radially
displaced fluid elements, radial advection of background flow,
magnetic tension, and the viscous force in the azimuthal direction.
The rate of change of the radial magnetic field perturbations is due
to the creation of radial field out of the vertical background frozen
into the radial velocity field with a vertical gradient.  Finally, the
rate of change of the azimuthal magnetic field perturbations is due to
the shearing of radial magnetic field perturbations.

\begin{figure*}[t]
  \includegraphics[width=0.675\columnwidth,trim=0 0 0 0]{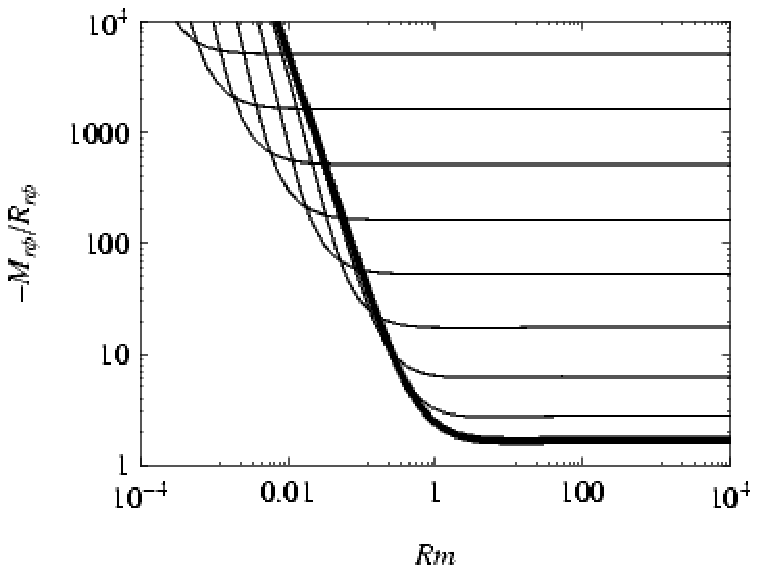}
  \includegraphics[width=0.675\columnwidth,trim=0 0 0 0]{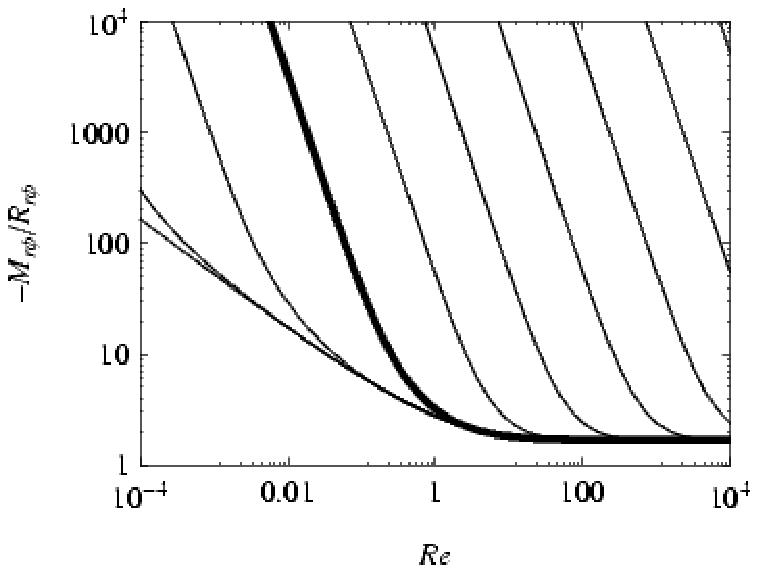}
  \includegraphics[width=0.675\columnwidth,trim=0 0 0 0]{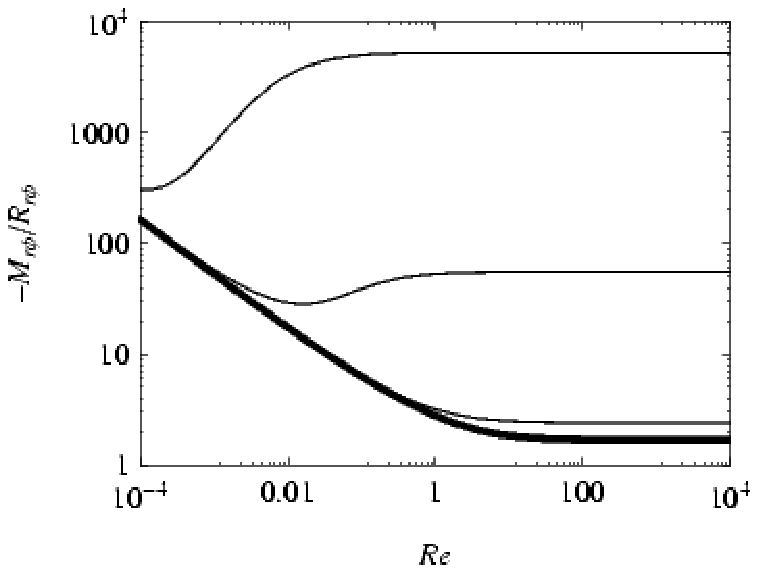}
  \caption{Ratio between the Maxwell and the Reynolds stresses,
    $-\bar{M}_{r\phi}/\bar{R}_{r\phi}$, for Keplerian rotation in
    different dissipative regimes.  \emph{Left}: Ratio between the
    Maxwell and the Reynolds stresses as a function of the magnetic
    Reynolds number for different values of the Reynolds number.  The
    thick solid line denotes the inviscid limit, i.e., ${\rm
      Re}\rightarrow \infty$.  The thin solid lines, in increasing
    order according to their asymptotic value at high magnetic
    Reynolds numbers correspond to ${\rm Re} =1, 0.1, \ldots$.  For
    ${\rm Rm}>1$, the ratio is independent of ${\rm Rm}$ regardless of
    the value of ${\rm Re}$.  \emph{Middle}: Ratio between the Maxwell
    and the Reynolds stresses as a function of the Reynolds number for
    different values of the magnetic Prandtl number. The various
    curves, from left to right, correspond to ${\rm Pm}=10^3, 10^2,
    \ldots$.  The thick solid line corresponds to the case ${\rm
      Pm}=1$.  \emph{Right}: Ratio between the Maxwell and the
    Reynolds stresses as a function of the Reynolds number for
    different values of the magnetic Reynolds number.  The thick solid
    line corresponds to the ideal conductor limit, i.e., ${\rm
      Rm}\rightarrow \infty$.  The thin solid lines, in increasing
    order according to their asymptotic value at high Reynolds numbers
    correspond to ${\rm Rm} = 1, 0.1, 0.01$. The Maxwell stress is
    larger than the Reynolds stress for any combination of the
    Reynolds and magnetic Reynolds numbers. The minimum value of this
    ratio is achieved in the ideal MHD regime, where
    $-\bar{M}_{r\phi}/\bar{R}_{r\phi} =5/3$, for Keplerian rotation.}
  \label{fig:stresses}
\end{figure*}

\section{Non-Ideal MRI-driven Stresses and Energy Densities}
\label{sec:stresses_energies}

The $i,j$ components of the mean Reynolds and Maxwell stresses
associated with the velocity and magnetic field perturbations are given
by
\begin{eqnarray}
\label{eq:mean_reynolds}
  \bar{R}_{ij}(t) & \equiv &  \langle \delta v_i(z,t) \, \delta  v_j(z,t) \rangle \, , \\ 
\label{eq:mean_maxwell}
  \bar{M}_{ij}(t) & \equiv &  \langle \delta b_i(z,t) \, \delta  b_j(z,t) \rangle \, ,
\end{eqnarray}
where the brackets denote mean values obtained via integration over
the disk scale-height, $2H$. These mean values can also be calculated
directly from the perturbations in Fourier space according to
\citep{PCP06}
\begin{eqnarray}
\label{eq:mean_reynolds_k}
\bar{R}_{ij}(t) & \equiv & 2 \sum_{n=1}^{\infty} \;
{\mathcal Re}[\,
\hat{\delta v_i}(k_n,t) \, \hat{\delta v_j}\!\!^*\!(k_n,t)\,] \,, \\
\label{eq:mean_maxwell_k}
\bar{M}_{ij}(t) & \equiv & 2 \sum_{n=1}^{\infty} \;
{\mathcal Re}[\,
\hat{\delta b_i}(k_n,t) \, \hat{\delta b_j}\!\!^*\!(k_n,t)\,] \,.
\end{eqnarray}
where ${\mathcal Re}[\,]$ stands for the real part of the quantity between square
brackets.  The off-diagonal components of these stresses relate to
angular momentum transport while their traces relate to the kinetic
and magnetic energy in the perturbations.

At late times, during the exponential growth of the instability, the
branch of unstable modes will dominate the growth of the perturbations
and the most important (secular) contribution to the mean stresses
will be given by the most unstable mode.  The leading order
contribution to these stress components are thus obtained by
considering the most unstable solutions to the set of equations
(\ref{eq:vx})--(\ref{eq:by}), which are given by equation
(\ref{eq:delta_zt_unstable}) when $k=k_{\rm max}$ with
$\gamma_+=\gamma_{\rm max}$.

\subsection{Non-ideal MRI Stresses}
\label{sub:MRI_stresses}

A measure of the angular momentum transport driven by the most
unstable MRI modes and mediated by the correlated perturbations in the
velocity and magnetic fields is obtained by setting
$(i,\,j)=(r,\,\phi)$ in equations (\ref{eq:mean_reynolds}) and
(\ref{eq:mean_maxwell}).  To leading order in time we obtain
\begin{eqnarray}
\label{eq:mean_Rrphi}
\bar{R}_{r\phi}(t) 
& = &
\frac{1}{2} \, 
\frac{v_0^2 e^{ 2\gamma_{\rm max} t}}{v_0^2+b_0^2} \, 
\sin(2\theta_{{\rm v}}) \,,\\
\label{eq:mean_Mrphi}
\bar{M}_{r\phi}(t) & = &
\frac{1}{2} \,
\frac{b_0^2 e^{ 2\gamma_{\rm max} t}}{v_0^2+b_0^2} \,
\sin(2\theta_{{\rm b}})\,. 
\end{eqnarray}

The results derived in \S~\ref{sec:mri_modes}, together with
Figures~\ref{fig:theta_diff},~\ref{fig:theta_mean},~and~\ref{fig:geom_sketch},
show that the angles $\theta_{\rm v}$ and $\theta_{\rm b}$
corresponding to the most unstable mode, $k_{\rm max}$, always satisfy
\begin{eqnarray}
  0 &\le& \theta_{\rm v} \le \arctan{\left(\frac{\kappa}{2}\right)} \le \frac{\pi}{4}\,, \\
 \frac{\pi}{2}   &\le& \theta_{\rm b} \le \frac{3\pi}{4} \,.
\end{eqnarray}
Both of these inequalities show explicitly that the mean Reynolds and
Maxwell stresses will be, respectively, positive and negative,
\begin{equation}
\bar{R}_{r\phi}(t) >0 \qquad \textrm{and} \qquad \bar{M}_{r\phi}(t) < 0 \,.
\end{equation}
This, in turn, implies that the mean total MRI-driven stress will be
always positive, i.e.,
\begin{equation}
\bar{T}_{r\phi}(t) = \bar{R}_{r\phi}(t) - \bar{M}_{r\phi}(t)  >0 \,,
\end{equation}
driving a net outward flux of angular momentum for any combination of
Reynolds and magnetic Reynolds numbers.

We conclude this section by calculating the ratio
$-\bar{M}_{r\phi}(t)/\bar{R}_{r\phi}(t)$ at late times
during the exponential growth of the instability.  
We obtain
\begin{equation}
\lim_{t\gg1}
\label{eq:analytical_ratio}
\frac{-\bar{M}_{r\phi}(t)}{\,\,\bar{R}_{r\phi}(t)} = -
\frac{v_0^2}{b_0^2} \frac{\sin(2\theta_{\rm v})}{\sin(2\theta_{\rm
    b})} \,.
\end{equation}
Using the definitions for the ratio $v_0/b_0$ (eq.~[\ref{eq:b0_v0}]) and
the angles $\theta_{\rm v}$ and $\theta_{\rm b}$
(eqs.~[\ref{eq:theta_v}] and [\ref{eq:theta_b}]), it can be seen that
the magnitude of the Maxwell stress, $-\bar{M}_{r\phi}(t)$, is always
larger than the magnitude of the Reynolds stress,
$\bar{R}_{r\phi}(t)$, provided that the flow is Rayleigh-stable, i.e.,
\begin{equation}
-\bar{M}_{r\phi}(t) > \bar{R}_{r\phi}(t)  \quad \textrm{for} \quad 0<q<2 \,.
\end{equation}
Figure~\ref{fig:stresses} shows the ratio between the Maxwell and the
Reynolds stresses in various dissipative regimes for Keplerian
rotation. Note that when the Reynolds and magnetic Reynolds numbers
are large enough we recover the result
$-\bar{M}_{r\phi}/\bar{R}_{r\phi}=5/3$, which coincides with the value
of this ratio in the ideal MHD case \citep{PCP06}.

\subsection{Non-ideal MRI Energetics}
\label{sub:MRI_energetics}

The mean energy densities associated with the perturbations in the
velocity and magnetic field are given by
\begin{eqnarray}
\label{eq:mean_EK}
\bar{E}_K(t) &=&
\frac{1}{2} \, (\bar{R}_{rr} + \bar{R}_{\phi\phi}) \,,\\
\label{eq:mean_EM}
\bar{E}_M(t) &=&
\frac{1}{2} \, (\bar{M}_{rr} + \bar{M}_{\phi\phi}) \,.
\end{eqnarray}
Substituting the expressions for the most unstable MRI-driven
perturbations from equation (\ref{eq:delta_zt_unstable}) into the
definitions for the diagonal components of the Reynolds and Maxwell
stresses, equations~(\ref{eq:mean_reynolds})--(\ref{eq:mean_maxwell}),
respectively, we obtain
\begin{eqnarray}
\label{eq:mean_R}
\bar{E}_K(t) 
& = & \frac{1}{2}\, \frac{v_0^2 e^{ 2\gamma_{\rm max} t}}{v_0^2+b_0^2} \,, \\
\label{eq:mean_M}
\bar{E}_M(t)
& = & \frac{1}{2}\, \frac{b_0^2 e^{ 2\gamma_{\rm max} t}}{v_0^2+b_0^2} \,.
\end{eqnarray}

\begin{figure*}[t]
  \includegraphics[width=0.675\columnwidth,trim=0 0 0 0]{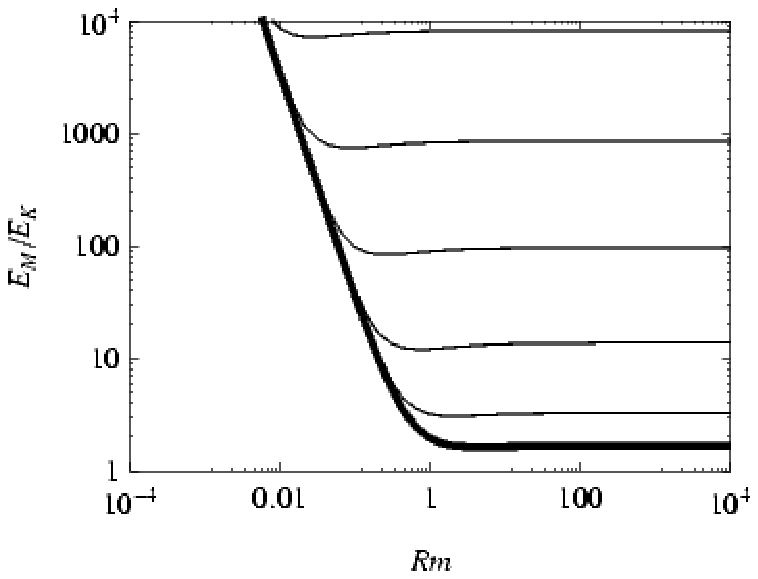}
  \includegraphics[width=0.675\columnwidth,trim=0 0 0 0]{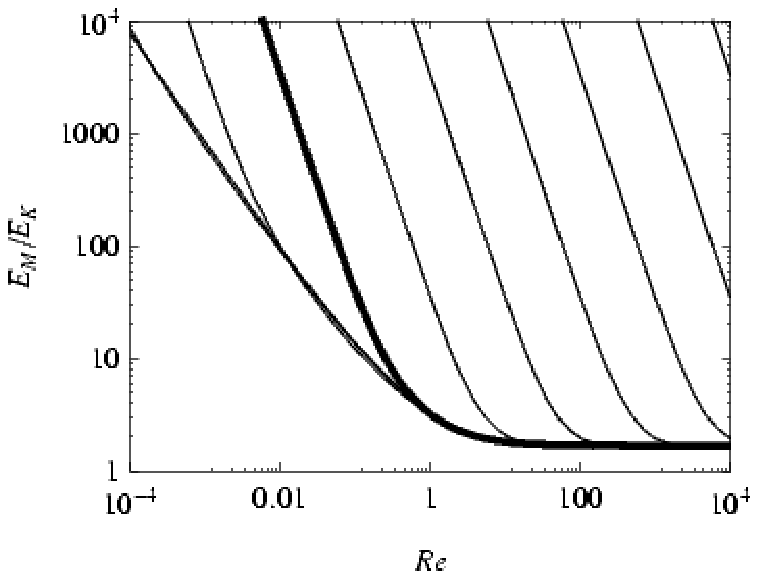}
  \includegraphics[width=0.675\columnwidth,trim=0 0 0 0]{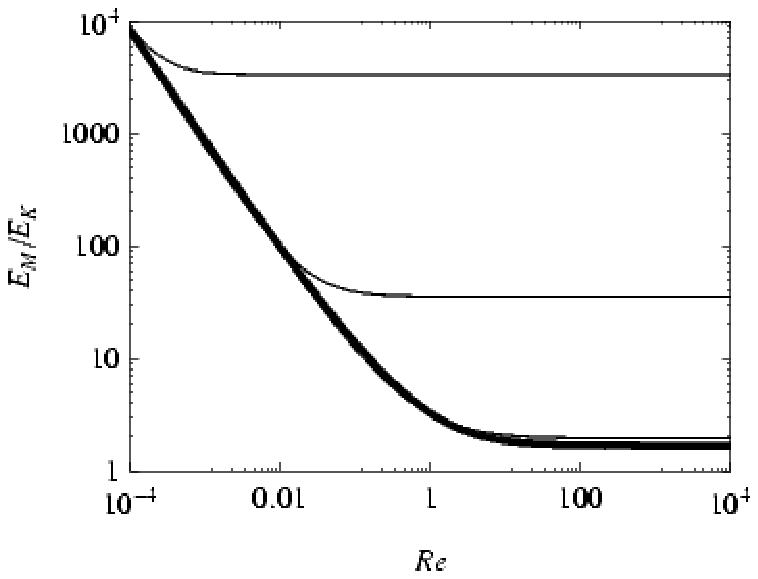}
  \caption{Ratio between the magnetic and the kinetic energy densities
    contained in MRI-driven perturbations, $\bar{E}_M/\bar{E}_K$, for
    Keplerian rotation in different dissipative regimes.  \emph{Left}:
    Ratio between the magnetic and the kinetic energy densities as a
    function of the magnetic Reynolds number for different values of
    the Reynolds number.  The thick solid line denotes the inviscid
    limit, i.e., ${\rm Re}\rightarrow \infty$.  The thin solid lines,
    in increasing order according to their asymptotic value at high
    magnetic Reynolds numbers correspond to ${\rm Re} = 1, 0.1,
    \ldots$.  For magnetic Reynolds numbers larger than unity, the
    ratio is independent of ${\rm Rm}$ regardless of the value of
    ${\rm Re}$.  Moreover, the asymptotic value of this ratio for
    ${\rm Rm}\ll 1$ is independent of the Reynolds number.
    \emph{Middle}: Ratio between the magnetic and the kinetic energy
    densities as a function of the Reynolds number for different
    values of the magnetic Prandtl number. The various curves, from
    left to right, correspond to ${\rm Pm}=10^2, 10, \ldots$.  The
    thick solid line corresponds to the case ${\rm Pm}=1$.
    \emph{Right}: Ratio between the Maxwell and the Reynolds stresses
    as a function of the Reynolds number for different values of the
    magnetic Reynolds number.  The thick solid line corresponds to the
    ideal conductor limit, i.e., ${\rm Rm}\rightarrow \infty$.  The
    thin solid lines, in increasing order according to their
    asymptotic value at high Reynolds numbers correspond to ${\rm Rm}
    = 1, 0.1, \ldots$.  Unlike the ratio between stresses, the ratio
    between energy densities seems to decrease monotonically with
    ${\rm Re}$ for any value of the magnetic Reynolds number. The
    magnetic energy density is larger than the kinetic energy density
    for any combination of the Reynolds and magnetic Reynolds numbers.
    The minimum value of this ratio is achieved in the ideal MHD
    regime, where $\bar{E}_M/\bar{E}_K=5/3$, for Keplerian rotation.}
  \label{fig:energies}
\end{figure*}

Using the definitions for the ratio $v_0/b_0$,
equation~(\ref{eq:b0_v0}), it can be seen that for non-ideal MRI modes
the mean energy associated with magnetic perturbations is always
larger than the mean energy corresponding to velocity perturbations as
long as the flow is Rayleigh-stable.  Figure~\ref{fig:energies} shows
the ratio between the magnetic and the kinetic energy densities in
various dissipative regimes for Keplerian rotation.

In the limit of large Reynolds and magnetic Reynolds numbers, we
recover the result $\bar{E}_M/\bar{E}_K=5/3$, which coincides with the
value of this ratio in the ideal MHD case and also with the ratio
between the magnitudes of the Maxwell and Reynolds stresses in the
ideal case \citep{PCP06}.  Note, however, that for arbitrary Reynolds
and magnetic Reynolds numbers, it is no longer true that the ratio
between mean magnetic and mean kinetic energies is equal to the ratio
between the magnitude of the mean Maxwell and the mean Reynolds
stresses. This can be seen by comparing
Figures~\ref{fig:stresses}~and~\ref{fig:energies}.

Finally, comparing equations (\ref{eq:mean_Rrphi}) and
(\ref{eq:mean_Mrphi}) with (\ref{eq:mean_R}) and (\ref{eq:mean_M}), it
immediately follows that
\begin{eqnarray}
 \bar{R}_{r\phi}(t) &\le& \bar{E}_K(t) \,,\\
-\bar{M}_{r\phi}(t) &\le& \bar{E}_M(t) \,.
\end{eqnarray}
This result, in turn, implies that the total mean energy associated
with the perturbations, $\bar{E}(t) = \bar{E}_K(t) + \bar{E}_M(t)$,
sets an upper bound on the total mean stress, i.e.,
\begin{eqnarray}
  \bar{T}_{r\phi}(t) \le \bar{E}(t) \,,
\end{eqnarray}
for any Reynolds and magnetic Reynolds numbers.

\section{Summary \& Discussion}
\label{sec:discussion}

We investigated the effects of viscosity and resistivity on the
stability of differentially rotating plasmas threaded by a magnetic
field perpendicular to the shear. We have shown that the most powerful
incompressible MRI modes are exact solutions of the MHD equations for
arbitrary combinations of the Reynolds and magnetic Reynolds numbers.
We have derived analytical expressions for the eigenfrequencies as
well as for the eigenmodes describing the MRI in viscous, resistive
media and provided a detailed description of the physical properties
of these modes in various dissipative regimes.

We have shown that the scalings derived for the marginally stable
mode, the most unstable wavenumber, and the maximum growth rate with
magnetic Reynolds number, $k_{\rm c}, k_{\rm max}, \gamma_{\rm max}
\propto {\rm Rm}$, valid for resistive, inviscid plasmas, see
equations~(\ref{eq:k_c_lim_reggrm}),~(\ref{eq:k_max_lim_reggrm})~and~(\ref{eq:gamma_max_lim_reggrm}),
as well as \citealt{SM99}, also hold when finite Reynolds numbers are
involved.  This is true as long as the magnetic Prandtl number is of
order unity or smaller, as it is usually the case in many
astrophysical systems (such as accretion disks around cataclysmic
variables and young stellar objects, as well as the Sun) and also in
MRI laboratory experiments.  Furthermore, we have addressed in detail,
for the first time to our knowledge, the physical properties of the
MRI in highly viscous, slightly resistive media.  These conditions are
expected to be found in the hot, diffuse gas in galaxies and galaxy
clusters. In this case, we found that the critical wavenumber for the
onset of the MRI, the most unstable wavenumber, and the maximum growth
rate scale with the Reynolds and magnetic Reynolds numbers according
to $k_{\rm c}\propto ({\rm Re\,Rm})^{1/3}$ and $k_{\rm max},
\gamma_{\rm max} \propto {\rm Re}^{1/2}$, see
equations~(\ref{eq:k_c_lim_rellrm}),~(\ref{eq:k_max_lim_rellrm})~and~(\ref{eq:gamma_max_lim_rellrm}).

We have provided a thorough geometrical description of the viscous,
resistive MRI modes in terms of the angles that define the planes
containing the velocity and magnetic field perturbations.  In the
ideal MHD limit, these planes are orthogonal, with the plane
containing the velocity disturbances laying at $45\degr$ with respect
to the radial direction.  We have shown that velocity and magnetic
field perturbations are still orthogonal if the magnetic Prandtl
number is unity, but that the planes containing them tend to be
aligned with the radial and azimuthal directions, respectively, when
the Reynolds number increases. In the regime of large Reynolds numbers
and small magnetic Reynolds numbers, magnetic and velocity field
perturbations tend to be orthogonal and aligned with the azimuthal and
radial directions, respectively. On the other hand, in the regime of
small Reynolds numbers and large magnetic Reynolds numbers, the
perturbed magnetic field tends to be aligned with the azimuthal
direction but the velocity field perturbations do not tend to be
aligned with the radial direction. The angle between both fields is
determined entirely by the epicyclic frequency $\kappa$. It would be
very interesting to understand to what extent this geometrical
dependence of MRI modes on the Reynolds and magnetic Reynolds numbers
influences the physical properties of kinetic and magnetic cells in
fully developed viscous, resistive MHD turbulence.

\begin{figure*}[t]
  \includegraphics[width=0.675\columnwidth,trim=0 0 0 0]{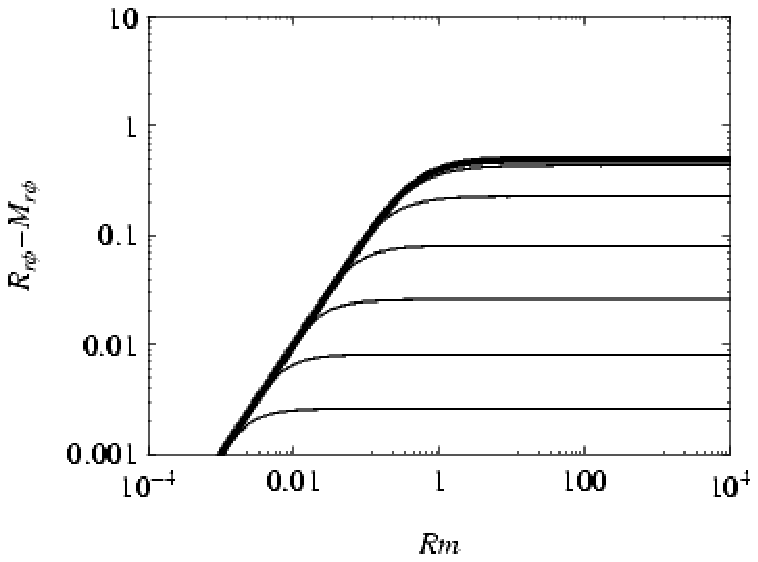}
  \includegraphics[width=0.675\columnwidth,trim=0 0 0 0]{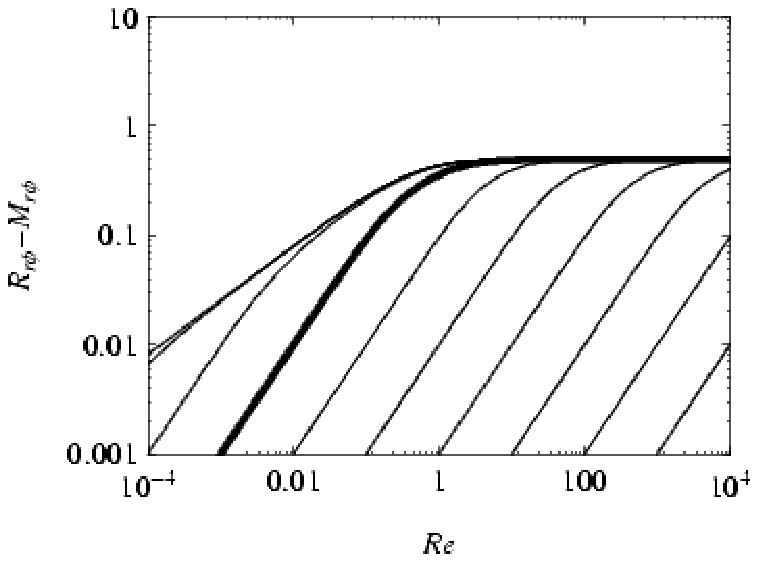}
  \includegraphics[width=0.675\columnwidth,trim=0 0 0 0]{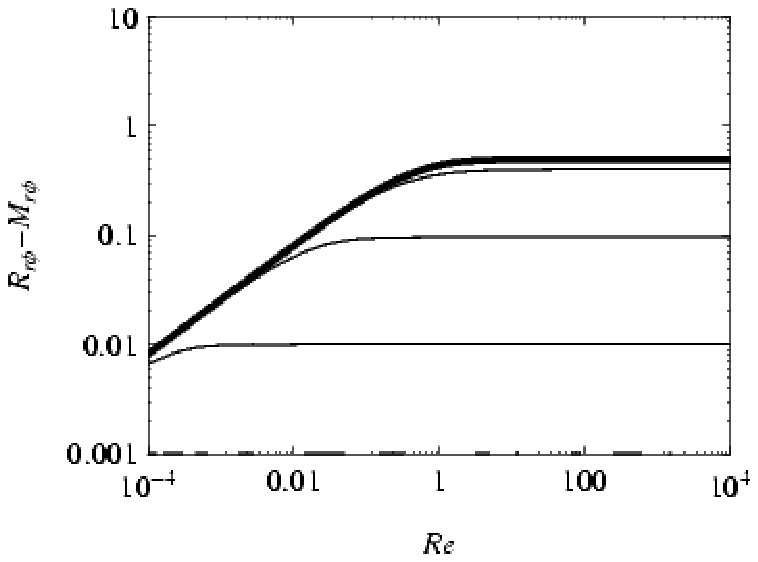}
  \caption{Mean total stress responsible for angular momentum
    transport, $\bar{T}_{r\phi}=\bar{R}_{r\phi}-\bar{M}_{r\phi}$,
    calculated according to
    equations~(\ref{eq:mean_Rrphi})~and~(\ref{eq:mean_Mrphi}), in
    various dissipative regimes for Keplerian rotation.  \emph{Left}:
    Mean total stress $\bar{T}_{r\phi}$ as a function of the magnetic
    Reynolds number for different values of the Reynolds number.  The
    thick solid line denotes the inviscid limit, i.e., ${\rm
      Re}\rightarrow \infty$.  The thin solid lines, in decreasing
    order, correspond to ${\rm Re} =10, 1, \ldots, 10^{-4}$.  For
    magnetic Reynolds numbers larger than unity, this ratio is
    independent of ${\rm Rm}$ regardless of the value of ${\rm Re}$.
    Moreover, the asymptotic value of this ratio for ${\rm Rm}\ll 1$
    is independent of the Reynolds number.  \emph{Middle}: Mean total
    stress $\bar{T}_{r\phi}$ as a function of the magnetic Prandtl
    number for different values of the Reynolds number.  From left to
    right, the curves correspond to ${\rm Pm}=10^3, 10^2, \ldots, 1$
    (thick solid line), $\ldots, 10^{-6}$.  \emph{Right}: Mean total
    stress $\bar{T}_{r\phi}$ as a function of the Reynolds number for
    different values of the magnetic Reynolds number.  The thick solid
    line corresponds to the ideal conductor limit, i.e., ${\rm
      Rm}\rightarrow \infty$. The thin solid lines, in decreasing
    order, correspond to ${\rm Rm} =10, 1, 0.1$.  Note that the stress
    scale in these plots is arbitrary.}
  \label{fig:Txy_RmPm}
\end{figure*}

In the ideal MHD limit, the exact (primary) MRI modes are known to be
unstable to parasitic (secondary) instabilities \citep{GX94}. These
parasitic modes have long been suspected to enable the mechanisms that
disrupt the primary modes providing an avenue toward saturation.  The
solutions derived in this work describe the dynamics of primary MRI
modes in viscous, resistive media enabling the study of parasitic
instabilities for arbitrary combinations of Reynolds and magnetic
Reynolds numbers.  The modifications in the relative directions of the
velocity and magnetic field perturbations characterizing the primary
viscous, resistive MRI modes described above can have an important
impact on the development and evolution of parasitic instabilities in
the presence of dissipation.

We have shown that, for any combination of the Reynolds and magnetic
Reynolds numbers, the mean Reynolds stress, $\bar{R}_{r\phi}=\langle
\delta v_r(z,t) \, \delta v_\phi(z,t) \rangle$, is always positive and
the mean Maxwell stress, $\bar{M}_{r\phi}=\langle \delta b_r(z,t) \,
\delta b_\phi(z,t) \rangle$, is always negative. This implies that the
mean total stress, $\bar{T}_{r\phi}=\bar{R}_{r\phi}-\bar{M}_{r\phi}$
is always positive, leading always to an outward transport of angular
momentum.  We have also demonstrated that both the ratio between
magnetic and kinetic stresses, $-\bar{M}_{r\phi}/\bar{R}_{r\phi}$, and
the ratio between magnetic and kinetic energy densities,
$\bar{E}_M/\bar{E}_K$, are always dominated by the magnetic
contribution.  These last two statements, support a somewhat
unexpected result since it is tempting to think that velocity
perturbations would dominate both the transport of angular momentum
and the energy density in highly resistive, inviscid plasmas.  It
would be very interesting to understand if and how the value of these
ratios in the saturated turbulent state vary as a function of the
Reynolds and magnetic Reynolds numbers.

Sano and collaborators have studied the linear \citep{SM99} and
non-linear \citep{SIM98, SI01, SS03, Sanoetal04} evolution of the MRI
for inviscid, resistive MHD.  The simulations in \citet{SS03} show
that for small magnetic Reynolds numbers the stresses at saturation
increase rapidly with increasing ${\rm Rm}$ and that there exists a
critical magnetic Reynolds number, of order unity, beyond which
turbulent stresses are rather insensitive to ${\rm Rm}$. The fact that
this same behavior is indeed seen when the stresses are due to
viscous, resistive MRI modes (see Figure~\ref{fig:Txy_RmPm}) rises the
question of how strong is the influence of long-lived, channel-like
modes on the fully developed turbulent state reached in shearing box
simulations with net magnetic flux through the vertical boundaries.

Systematic numerical studies of viscous, resistive MHD shearing flows
have begun to uncover the dependencies of microphysical dissipation on
the mean transport properties of MRI-driven turbulence. Numerical
simulations with both zero \citep{FPII07} and non-zero net magnetic
fluxes \citep{LL07} lead to the conclusion that angular momentum
transport increases with increasing magnetic Prandtl number when the
Reynolds number is held constant. This behavior can also be identified
when examining the stresses due to viscous, resistive MRI modes,
see~Figure~\ref{fig:Txy_RmPm}.

The effects of varying the Reynolds number at fixed magnetic Prandtl
number on the saturation of MRI-driven turbulence are currently rather
uncertain \citep[see, in particular, the discussion in][]{LL07}.  The
global trends exhibited by the available simulations suggest that the
stresses at saturation increase with increasing Reynolds number for
magnetic Prandtl numbers smaller than unity while the opposite
behavior is observed for magnetic Prandtl numbers larger than unity.
If confirmed, these results suggest that the mechanisms leading to
saturation might operate differently depending on whether the magnetic
Prandtl number is larger or smaller than unity.  In any case, having
obtained a better understanding of the behavior of the most unstable
MRI modes as a function the magnetic Prandtl number it would be very
interesting to follow the evolution of the viscous, resistive MRI from
the linear to the non-linear regime. By performing numerical
simulations with the same Prandtl number (both larger and smaller than
unity) and different Reynolds numbers we could see whether there is an
inversion of the trends observed in the linear regime (i.e., higher
stresses at higher Reynolds numbers for fixed magnetic Prandtl
numbers) after the exact solutions break down.  The comparison between
the late time behavior of the viscous, resistive MRI modes and fully
developed MHD turbulence with dissipation will shed light into the
mechanisms that lead to the saturation of the MRI in different
dissipative regimes.

Finally, most current numerical algorithms employ finite difference
methods (with constrained transport for the evolution of the magnetic
field).  The leading order errors in first-order upwind methods behave
like diffusion, however, in second-order central difference methods,
the leading order errors are dispersive (artificial viscosities are
usually employed to damp unphysical oscillations near shocks). The
comparison between numerical solutions from different algorithms can
provide estimates of these errors.  However, it is difficult to
quantify these numerical artifacts based on analytical studies of
ideal MHD. The analytical solutions derived in this paper, on the
other hand, describe the effects of arbitrary combinations of
viscosity and resistivity [see also \citet{LB07} who derived results
to leading order in $(\eta-\nu)/k$]. It should now be possible to
better measure the numerical viscosity and resistivity for a wide
range of Reynolds and magnetic Reynolds numbers by comparing numerical
solutions and analytical solutions of non-ideal MRI.  The results
presented in this paper provide ideal benchmarks to the study
numerical artifacts generated by different algorithms in various
dissipative regimes.

\appendix

\section{Analytical Solution to Quartic Equation and Ideal MHD Limit}
\label{sec:appendix}

The solutions to a depressed quartic equation of the form 
\begin{equation}
  \sigma^4 + \alpha \sigma^2 + \beta \sigma + \lambda = 0.
  \label{eq:quartic}
\end{equation}
are given by\footnote{Note that the two $\pm_b$'s have the same sign
  so there are only four solutions instead of eight.}
\begin{equation}
  \sigma =   \pm_a \sqrt{-\left(\frac{3\alpha}{4} + \frac{y}{2} \pm_b
      \frac{\beta/4}{\sqrt{\alpha/4 + y/2}}\right)} \pm_b \sqrt{\frac{\alpha}{4} + \frac{y}{2}}  \,,
  \label{eq:solution_sigma}
\end{equation}
where $y$ is any of the solutions of the cubic equation
\begin{equation}
\label{eq:cubic}
  y^3 + \frac{5\alpha}{2} y^2 + (2\alpha^2 - \lambda) y +
  \left(\frac{\alpha^3}{2} - \frac{\alpha\lambda}{2} -
  \frac{\beta^2}{8}\right) = 0 \,.
\end{equation}

In the special case $\beta \rightarrow 0$, the solutions to equation
(\ref{eq:cubic}) take simple forms.  In order to see how the general
solutions reduce to the simple cases, we write the cubic equation as
\begin{equation}
  \left(y + \frac{\alpha}{2}\right)\left[(y + \alpha)^2 -
    \lambda\right] = \frac{\beta^2}{8} \,.
\end{equation}
It is easy to see that, if $y \ne -\alpha/2$,
\begin{equation}
  \frac{\beta/4}{\sqrt{\alpha/4 + y/2}} = \sqrt{(y + \alpha)^2 - \lambda}.
\end{equation}
Using the above identity, the general
solution~(\ref{eq:solution_sigma}) becomes
\begin{equation}
  \sigma = \pm_a
\sqrt{-\Lambda \pm_b \sqrt{\Delta}} 
\pm_b \, \frac{\beta/4}{\sqrt{\Delta}} \,,
\end{equation}
where we have defined
\begin{equation}
  \Lambda = \frac{3\alpha}{4} + \frac{y}{2} 
  \qquad \textrm{and} \qquad
  \Delta = \sqrt{(y + \alpha)^2 - \lambda} \,.
\end{equation}

If we choose the root so that
\begin{equation}
  \lim_{\beta \rightarrow 0} y = -\frac{\alpha}{2},
\end{equation}
it is easier to take the limit 
\begin{equation}
\lim_{\beta \rightarrow 0} \sigma = 
\lim_{\beta \rightarrow 0} \pm \sqrt{-\Lambda \pm \sqrt{\Delta}} =
  \pm \sqrt{- \frac{\alpha}{2} \mp \sqrt{\frac{\alpha^2}{4} - \lambda}}
  \equiv \pm \sqrt{-\Lambda_0\pm\sqrt{\Delta_0}} = \sigma_0
\end{equation}
where $\Lambda_0$ and $\Delta_0$ are defined in equations
(\ref{eq:Lambda0}) and (\ref{eq:Delta0}).

\acknowledgments{We thank Jeremy Goodman and Dimitrios Psaltis for
  valuable comments and discussions. We are grateful to Roman
  Shcherbakov for the initial discussions that lead to the idea of
  combining the dispersion relation and its derivative to obtain the
  asymptotic expressions derived in \S~\ref{sec:mri_modes}.  MEP
  gratefully acknowledges support from the Institute for Advanced
  Study. CKC is supported through an ITC Fellowship at Harvard. MEP and
  CKC are grateful to the Harvard-Smithsonian Institute for Theory and
  Computation and the Institute for Advanced Study, respectively, for
  their hospitality during part of this work.}

%%%%%%%%%%%%%%%%%%%%%%%%%%%%%%%%%%%%%%%%%%%%%%%%%%%%%%%%%%%%%%%%%%%%%%%


\begin{thebibliography}{}

\bibitem[Balbus \& Hawley(1991)]{BH91}
{Balbus, S.\ A. \& Hawley J.\ F. 1991, ApJ, 376, 214}

\bibitem[Balbus \& Hawley(1992)]{BH92}
{Balbus, S.\ A. \& Hawley J.\ F. 1992, ApJ, 392, 662}

\bibitem[Balbus \& Hawley(1998)]{BH98}
{---------. 1998, Rev. Mod. Phys., 70, 1}

\bibitem[Balbus \& Henri(2007)]{BH08}
{Balbus, S.\ A. \&  Henri, P. 2008, ApJ, in press [arXiv0706.0828]}

\bibitem[Blaes \& Balbus(1994)]{BB94}
{Blaes, O.\ M., \& Balbus, S.\ A. 1994, ApJ, 421, 163}

\bibitem[Brandenburg {et~al.}(1995) Brandenburg, Nordlund, Stein, \& Torkelsson]{Brandenburg95} 
{Brandenburg, A., Nordlund, A., Stein, R.\ F., \& Torkelsson, U. 1995, ApJ, 446, 741}

\bibitem[Fleming, Stone, \& Hawley(2000)]{FSH00}
{Fleming, T.\ P., Stone, J.\ M., Hawley 2000, ApJ, 530, 464}

\bibitem[Fromang \& Papaloizou(2007)]{FPI07}
{Fromang, S. \& Papaloizou, J. 2007, A\&A 476, 1113}

\bibitem[Fromang {et~al.}(2007)Papaloizou, Lesur, \& Heinemann]{FPII07}
  {Fromang, S., Papaloizou, J., Lesur, G., \& Heinemann, T. 2007,
    A\&A, 476, 1123}

\bibitem[Gammie(1996)]{Gammie96}
{Gammie, C.\ F. 1996, ApJ, 462, 725}	

\bibitem[Goodman \& Xu(1994)]{GX94}
{Goodman, J. \& Xu, G. 1994, ApJ, 432, 213}

\bibitem[Goodman \& Ji(2002)]{GJ02}
{Goodman, J. \& Ji, H. 2002, JFM, 462, 365}	

\bibitem[Hawley, Gammie, \& Balbus(1995)]{HGB95}
{Hawley, J.\ F., Gammie, C.\ F., \& Balbus, S.\ A. 1995, ApJ, 440, 742}

\bibitem[Hoffman \& Kunze(1971)]{HK71}
{Hoffman K.M., Kunze R., 1971, Linear Algebra, 2nd ed. Prentice Hall, N.J.}

\bibitem[Ji, Goodman \& Kageyama(2001)]{JGK01}
{Ji, H., Goodman, J. \&  Kageyama, A. 2001, MNRAS, 325, L1}

\bibitem[Jin(1996)]{Jin96}
{Jin, L. 1996, ApJ, 457, 798}

\bibitem[Liu, Goodman, \& Ji(2006)]{LGJ06}
{Liu, W., Goodman, J. \& Ji, H. 2006, ApJ, 643, 306}

\bibitem[Lesur \& Longaretti(2007)]{LL07}
{Lesur, G. \& Longaretti, P.\ Y. 2007, MNRAS, 378, 1471}

\bibitem[Lesaffre \& Balbus(2007)]{LB07}
{Lesaffre, P. \& Balbus, S.\ A. 2007, MNRAS, 381, 319}

\bibitem[Pessah \& Psaltis(2005)]{PP05}
{Pessah, M.\ E. \& Psaltis, D. 2005, ApJ, 628, 879}

\bibitem[Pessah, Chan, \& Psaltis(2006)]{PCP06}
{Pessah, M.\ E., Chan C.\ K., \& Psaltis, D. 2006, MNRAS, 372, 183}

\bibitem[Pessah, Chan, \& Psaltis(2007)]{PCP07}
{Pessah, M.\ E., Chan C.\ K., \& Psaltis, D. 2007, ApJ, 668, L51}

\bibitem[R\"udiger, Schultz, Shalybkov(2003)]{SSW03} 
{R\"udiger, G., Schultz, M., \& Shalybkov, D. 2003, Phys. Rev. E, 67, 046312}

\bibitem[Salmeron \& Wardle(2005)]{SW05}
{Salmeron, R. \& Wardle, M. 2005, MNRAS, 361, 45}	

\bibitem[Sano \& Inutsuka(2001)]{SI01}
{Sano, T. \&  Inutsuka, S.\ I. 2001, ApJ, 561, L179}	

\bibitem[Sano, Inutsuka, \& Miyama(1998)]{SIM98}
{Sano, T., Inutsuka, S.\ I., \& Miyama, S.\ M. 1998, ApJ, 506, L57}

\bibitem[Sano {et~al.}(2004)Sano, Inutsuka, Turner, \& Stone]{Sanoetal04} 
{Sano, T., Inutsuka, S.\ I., Turner, N.\ J., \& Stone, J.\ M. 2004, ApJ, 605, 321}

\bibitem[Sano \& Miyama(1999)]{SM99}
{Sano, T. \& Miyama, S.\ M. 1999, ApJ, 515, 776}

\bibitem[Sano \& Stone(2003)]{SS03} {Sano, T. \& Stone, J.\ M. 2003,
    in Scientific Frontiers in Research on Extrasolar Planets, ASPC,
    Vol 294, D., Deming \& S., Seager, eds. (San Francisco: ASP)}

\bibitem[Sisan {et~al.}(2004)Sisan, Mujica, Tillotson, Huang, Dorland,
  Hassam, Antonsen, \& Lathrop]{Sisanetal04} 
{Sisan, D.\ R. Mujica, N., Tillotson, W.\ A., Huang, Y.\ M.,
Dorland, W., Hassam, A.\ B., Antonsen, T.\ M., \& Lathrop, D.\ P.
2004, Phys.  Rev. Lett., 93, 114502}

\bibitem[Turner, Sano, \& Dziourkevitch(2007)]{TSD07}
{Turner, N.\ J., Sano, T. \& Dziourkevitch, N. 2007, ApJ, 659, 729}

\end{thebibliography}
\end{document}